\documentclass[useAMS,usenatbib]{mn2e}
\usepackage{psfig}
\usepackage{amssymb}
\usepackage{color}
\usepackage{enumerate}
\usepackage{rotating}
\usepackage{amsmath}
\usepackage{mathrsfs}
\usepackage{ragged2e}

\title[High-Mass End of the L-$\sigma$ relation at $z\sim0.55$]
{A Steep Slope and Small Scatter for the High-Mass End of the L-$\sigma$ Relation at $z\sim0.55$}

\author[Montero-Dorta et al.]{
\parbox[t]{\textwidth}{
Antonio D. Montero-Dorta$^{1}$\thanks{E-mail: amontero@astro.utah.edu}, Yiping Shu$^{1}$, Adam S. Bolton$^{1}$, Joel R. Brownstein$^{1}$, Benjamin J. Weiner$^{2}$}
\vspace*{6pt} \\ 
$^1$ Department of Physics and Astronomy, The University of Utah, 115
South 1400 East, Salt Lake City, UT 84112, USA \\
$^2$ Steward Observatory, 933 N. Cherry Ave., University of Arizona, Tucson, AZ 85721, USA \\
\vspace{-0.5cm} 
}

\date{Accepted ---. Received ---;in original form ---}

\def\simlt{\lower.5ex\hbox{$\; \buildrel < \over \sim \;$}}
\def\simgt{\lower.5ex\hbox{$\; \buildrel > \over \sim \;$}}
\usepackage{graphicx}
\usepackage{rotating}

\definecolor{red}{rgb}{1,0,0}

\begin{document}

\bibliographystyle{mnras}

\maketitle

\begin{abstract}

We measure the intrinsic relation between velocity dispersion ($\sigma$) and 
luminosity ($L$) for massive, luminous red galaxies (LRGs) at redshift $z \sim 0.55$. We achieve 
unprecedented precision by using a sample of 600,000 galaxies with spectra from the Baryon 
Oscillation Spectroscopic Survey (BOSS) of the third Sloan Digital Sky Survey (SDSS-III), covering 
a range of stellar masses $M_* \gtrsim 10^{11} M_{\odot}$. We deconvolve the effects of photometric errors, limited spectroscopic signal-to-noise ratio, and red--blue 
galaxy confusion using a novel hierarchical Bayesian formalism that is generally applicable to any 
combination of photometric and spectroscopic observables. For an L-$\sigma$ relation of the form 
$L \propto \sigma^{\beta}$, we find $\beta = 7.8 \pm 1.1$ for $\sigma$ corrected to the effective radius, and a very small intrinsic scatter of $s = 0.047 \pm 0.004$ 
in $\log_{10} \sigma$ at fixed $L$. No significant redshift evolution is found for these parameters. The evolution of the zero-point within the redshift range considered is consistent with the passive evolution of a galaxy population that formed at redshift $z=2-3$, assuming 
single stellar populations. An analysis of previously reported results seems to indicate that 
the passively-evolved high-mass L-$\sigma$ relation at $z\sim0.55$ is consistent with the one measured at $z=0.1$.
Our results, in combination with those presented in \cite{MonteroDorta2014}, provide a detailed description of the high-mass end of the red sequence (RS) at $z\sim0.55$. This characterization, in the light of previous literature, suggest that the high-mass RS distribution corresponds to the ``core" elliptical population.

\end{abstract}


\section{Introduction}
\label{sec:intro}

In the 60's and 70's, several empirical scaling relations between the kinematic and photometric properties of early-type galaxies (ETGs) were
identified. Thanks to the seminal work of  \cite{Djorgovski1987} and  \cite{Dressler1987}, today we know that these relations are different projections of the so-called 
\emph{fundamental plane} (FP) of ETGs, which is a thin plane outlined by the occupation of ETGs in the
three-dimensional space spanned by velocity dispersion, effective radius and surface brightness, i.e., 
$\log_{10} \sigma$, $\log_{10} R_e$, and $\log_{10} \langle I \rangle_e$, respectively. 

Of particular interest among these scaling relations is the L-$\sigma$ relation, a specific two-dimensional projection of the FP that 
relates the luminosity $L$ and the central stellar velocity dispersion $\sigma$ of ETGs.  This relation was first
reported by \cite{Minkowski1962}, using a sample of only 13 ETGs, and by \cite{Morton1973} a decade later, although
no quantification was provided. The relation was first quantified in a sample of 25 galaxies by \cite{FJR}, as a power law in 
the form $L \propto \sigma^4$, and has been commonly called the \emph{Faber-Jackson Relation} (F-J relation) since then. 
As a link between a distance-independent quantity $\sigma$ and an intrinsic property $L$, the F-J relation immediately became a useful distance estimator 
and consequently a cosmological probe \citep[e.g.,][]{deVaucouleurs1982a, deVaucouleurs1982b, Pature1992}. 

With the emergence of large-scale structure (LSS) galaxy surveys in the last decades, the size of the ETG samples
available increased dramatically, especially at low redshift, i.e. $z\sim0.1$ (mostly with the 
Sloan Digital Sky Server, SDSS, \citealt{York2000}). It was then confirmed with 
statistical significance, that the L-$\sigma$ relation at intermediate masses/luminosities approximately 
follows a canonical F-J relation with a slope of $\sim4$ (note that the words ``slope" an ``exponent"
are commonly used interchangeably, as the relation is often expressed in the form $M \propto \log_{10} \sigma$, where M is the absolute magnitude).
This result is reported in, e.g., \cite{Bernardi2003b} and \cite{Desroches2007}, 
along with a typical scatter in $\log_{10} \sigma$ of $\sim0.1$ dex. 

Almost since the very discovery of the L-$\sigma$ relation, however, it became clear that the slope
depended on the luminosity range of the sample under analysis. A value of $\gtrsim 4.5$ has been measured 
in luminous ETGs \citep[e.g.,][]{Schechter1980, Malumuth1981, Cappellari2013_XX, Kormendy2013} and 
$\sim 2$ in faint ETGs \citep[e.g.,][]{Tonry1981,Davies1983, Held1992, deRijcke2005, Matkovic2007}. 
A confirmation of the curvature of the L-$\sigma$ relation towards the high-mass end with high statistical 
significance has been possible thanks to the SDSS (see \citealt{Desroches2007, Hyde2009a, NigocheNetro2010, Bernardi2011}).
The $z\sim0.1$ sample of \citealt{Bernardi2011}, as an example, contains $\sim18,000$ ETGs 
with $\log_{10} M_* \gtrsim 11.2/11.5$. As the F-J relation, where $L \propto \sigma^4$, is therefore a particular, canonical case that only appears to hold
for morphologically selected samples at intermediate mass ranges, 
in the remainder of this paper we adopt the generic ``L-$\sigma$ relation" terminology.

 The curvature of the L-$\sigma$ relation is a consequence of the curvature of the FP itself. In fact, the 
phenomenology of the FP has turned out to be rather complex. Enough evidence has been gathered that its characteristics 
depend, not only on luminosity/stellar mass but, to a greater or lesser degree, on a variety of other galaxy properties. 
In addition, the sensitivity of the measurements to selection effects and low-number 
statistics has often lead to contradictory results. In terms of redshift evolution, it appears clear that the zero-point of the FP must evolve from $z\sim1$
in a way that approximates that of a passively-evolving galaxy population \citep[e.g.,][]{Bernardi2003c,Jorgensen2006}. 
However, there are some indications that the FP might be significantly steeper at higher redshifts \citep{Jorgensen2006, Fritz2009}, 
whereas for the intrinsic scatter, conclusions are so far somewhat contradictory  (e.g., \citealt{NigocheNetro2011} measure a decrease in the scatter of the 
the L-$\sigma$ relation, whereas results from \citealt{Shu2012} appear to indicate the opposite). 
It seems that the properties of the FP also depend, to some extent, on the
wavelength range probed, which is an indication that stellar population properties have an impact on the FP. 
While these properties appear to remain fairly unchanged in the optical \citep{Bernardi2003c,Hyde2009b}, there are clear indications that the FP is intrinsically different in the infrared \citep[e.g.,][]{Pahre1998,LaBarbera2010,Magoulas2012}. 
The dependence of the FP on environment has also been the subject of a number of works (see e.g. \citealt{Treu2001,Bernardi2003c, Reda2004,Reda2005,Denicolo2005,LaBarbera2010}, 
for the FP;  \citealt{Focardi2012} for the L-$\sigma$ relation). This 
is not by any means a complete list; a number of other galaxy properties have been investigated in recent years.
 
 While the use of the FP and its projections as a cosmological probe has been largely eclipsed by other 
techniques, the FP has gained increasing attention in the last decades as a source of observational constraints
for galaxy formation and evolution theories. In this sense, the variation and evolution trends with respect to galaxy properties in the slope and intrinsic scatter of the FP and it projections are believed to be the imprints of non-homological physical processes that occur during the formation and evolution of galaxies. Understanding these observed trends provides crucial insight into these fundamental 
processes. In this sense, recent cosmological simulations have investigated the impacts of various physical 
processes such as major/minor mergers and disc instabilities on shaping the internal structure and 
kinematic properties of ETGs \citep[e.g.,][]{Oser2012, Shankar2013, Posti2014, Porter2014}.

 In this paper, we use data from the Baryon Oscillation Spectroscopic Survey (BOSS, \citealt{Dawson2013}) of the SDSS-III 
\citep{Eisenstein2011} to measure, for the first time, 
the high-mass end of the L-$\sigma$ relation at $0.5<z<0.7$. This is a continuation 
of the work presented in \cite{MonteroDorta2014}, hereafter MD2014, where the intrinsic colour-colour-magnitude red sequence (RS) 
distribution is deconvolved from photometric errors and selection effects in order to compute the evolution 
of the RS luminosity function (LF). An important conclusion of MD2014 is that, at fixed apparent magnitude, and 
for a narrow redshift slice, the RS is an extremely narrow distribution ($<0.05$ mag), consistent with a single point in the optical 
colour-colour plane. This work is intended to measure the L-$\sigma$ relation that this photometrically 
distinct population obeys. At intermediate redshifts, the BOSS capability to characterize the massive RS population is
unrivaled, with a huge sample of more than 1 million luminous red galaxies (LRGs) with stellar masses $M_* \gtrsim 10^{11} M_{\odot}$. 
No other previous survey or sample has been able to probe this population with comparable statistics, hence 
the unique value of the measurements reported here (find a preliminary analysis from BOSS in \citealt{Shu2012}, where incompleteness is, however, only partially addressed).
Importantly, our intrinsic RS was identified in MD2014 using exclusively photometric information, i.e. our red-blue
deconvolution is based on the phenomenology of the colour-colour plane, and not on any morphological 
classification. We will, therefore, use the ``RS" terminology instead of the ``ETG'' terminology when referring to our sample/results. 

The mass range covered by BOSS has been hard to probe at $z\sim0.55$. In fact, most of the information that 
we have about the high-mass end of the L-$\sigma$ relation comes from the SDSS at $z\sim0.1$ or 
from small samples of very-nearby ETGs, using high-resolution observations (the latter being strongly 
affected by low-number statistics and selection effects). One of the most important discoveries from these studies is that 
the slope of the L-$\sigma$ relation is steeper at higher masses/luminosities, as mentioned above. 
In recent years, a picture that attempts to explain this mass dependence has emerged. The curvature in 
the scaling relations has been associated with a characteristic stellar mass scale of $\sim 2 \times 10^{11} M_{\odot}$. This 
scale, which was first reported at high significance by \cite{Bernardi2011} with the SDSS, is thought to be related to a 
change in the assembly history of galaxies (recently, \citealt{Bernardi2014} have shown that this 
scale is also special for the late-type galaxy population).

High-resolution imaging has shown that the high-mass scale marks the separation between  
two distinct ETG populations: {\it{core ellipticals}} are defined by the fact that the central light 
profile is a shallow power law separated by a break from the outer, steep Sersic function profile, whereas 
in {\it{coreless ellipticals}} (also known as ``power-law" or ``cusp" ellipticals) this feature is not present (see e.g. \citealt{Lauer2007a, Lauer2007b}).
It has been shown in small samples that core ellipticals dominate 
at the high-mass end, while coreless ellipticals are predominant at lower masses 
\citep{Kormendy1996, Faber1997,Hyde2008,Cappellari2013_XX,Kormendy2013}.
Importantly, this bimodality in the central surface brightness profile extends to a variety of other 
properties. The distinct characteristics of each of these types, including the fact that core ellipticals obey an L-$\sigma$ relation
with a significantly steeper slope \citep{Lauer2007a, Kormendy2013}, have been associated with 2 different evolutionary 
paths for these objects. Core ellipticals are thought to be formed through 
major dissipation-less mergers \citep{Desroches2007, vonderLinden2007, Hyde2008, Lauer2007b, Bernardi2011,Cappellari2013_XX, Kormendy2013},
whereas coreless ellipticals might have undergone more recent episodes of star formation (\citealt{Kormendy2009} review evidence that they are formed in wet mergers 
with starbursts). 

To complement the work done in MD2014, here we present a novel method to combine
photometric and spectroscopic quantities in low signal-to-noise (SN) large-photometric-error samples that 
we call {\it{Photometric Deconvolution of Spectroscopic Observables}} (hereafter, PDSO).
The PDSO is a hierarchical Bayesian statistical method that allows us to combine the velocity dispersion likelihood function measurements of  \cite{Shu2012} with 
the photometric red/blue deconvolution of MD2014 to provide the most precise measurement ever performed of the high-mass end of the intrinsic L-$\sigma$ relation within the redshift range $0.5<z<0.7$. 

This paper is organized as follows. Section~\label{sec:overview} provides an overview of methods and motivations. In Section~\ref{sec:data} we briefly describe 
the target selection for the galaxy sample that we use, the BOSS CMASS sample (\ref{sec:cmass}), and the computation of stellar velocity dispersion 
likelihood functions from \cite{Shu2012} (\ref{sec:vdisp}). Section~\ref{sec:intrinsic} is devoted to summarizing the results of MD2014 
regarding the intrinsic RS colour-colour-magnitude distribution. In Section~\ref{sec:formalism0}, we
present our PDSO method n a general form (\ref{sec:formalism}) and 
we addresses the application of our method to the BOSS CMASS sample (\ref{sec:application}). In
Section~\ref{sec:aperture}, we describe our aperture correction procedure. In Section~\ref{sec:results} 
we present the best-fit parameters for the $\sigma$ -- apparent magnitude relation (\ref{sec:parameters}),
discuss the effect of addressing completeness and the red/blue population deconvolution (\ref{sec:effect})
and present our the L-$\sigma$ relation results (\ref{sec:F-J relation}). In Section~\ref{sec:discussion}, we compare our measurements 
with previous results from the literature (\ref{sec:comparison}) and discuss
on the physical implications of our measurements (\ref{sec:interpretation}). Finally, we summarize our main conclusions and discuss future applications in Section~\ref{sec:conclusions}. 
Throughout this paper we adopt a cosmology with $\Omega_M=0.274$,  $\Omega_\Lambda=0.726$ and $H_0 = 100h$ km s$^{-1}$ Mpc$^{-1}$ with $h=0.70$ 
(WMAP7, \citealt{Komatsu2011}), and use AB magnitudes \citep{OkeGunn1983}.

\section{Overview of methods and motivations}
\label{sec:overview}

 The statistical power of BOSS to cover the very-massive RS population is unrivaled at $z\sim 0.55$, with 
a sample of $\sim1$ million galaxies (in the latest data release, see \citealt{Alam2015}) with stellar masses $M_* \gtrsim 10^{11} M_{\odot}$ and a median stellar mass 
of $M_* \simeq 10^{11.3} M_{\odot}$ (as measured by \citealt{Maraston2013}, assuming a \cite{Kroupa2001}
initial mass function). The samples, however, present significant challenges, including low SN ratio
for the spectra, large photometric errors and a selection scheme that allows for a fraction of bluer objects that increases
with redshift. The photometric issues are addressed in MD2014, where we photometrically deconvolve the intrinsic 
red sequence distribution from photometric errors and selections effects. Our red--blue population deconvolution 
allows us to characterize completeness in the CMASS sample, which is the main LRG sample from BOSS, covering a redshift range 
$0.4<z<0.7$ (see next section for a complete description of the sample). This characterization allows us to analyze the luminosity function 
and colour evolution of the LRG population. 

 The aforementioned analysis in MD2014 is performed within the framework of a Bayesian hierarchal statistical method that 
is aimed at constraining distributions of galaxy properties, instead of individual-galaxy properties (since 
an object-by-object approach is discouraged by the characteristics of the BOSS data). The same 
philosophy is applied in this work to compute the high-mass end of the L-$\sigma$ relation. The main
steps of our analysis are:

\begin{itemize}
	\item Development of a general Bayesian hierarchal statistical method that combines the photometric red-blue deconvolution and selection function from
	MD2014 with probability density information from a spectroscopic observable. We call this formalism
	{\it{photometric deconvolution of spectroscopic observables}} or PDSO. Our method is aimed at constraining the 
	hyper-parameters of a model for the joint pdf of survey galaxies in physical parameter space, 
	by marginalizing over the physical parameter likelihood functions of individual galaxies given the survey data. 
	\item Application of the above formalism to the computation of the best-fit intrinsic L-$\sigma$ relation from BOSS. The idea is
	to parametrize the intrinsic distribution in L-$\sigma$ space, and use the PDSO method to constrain the parameters of this distribution:
	the slope, zero-point and the intrinsic scatter of the L-$\sigma$ relation.
	\item Evaluation of the redshift evolution of the best-fit parameters that define the L-$\sigma$ relation within a suitable redshift range, namely $0.5<z<0.7$.	
\end{itemize}

 The results of the above analysis will add important constraints to the 
evolution of massive RS galaxies. In addition, the PDSO method will 
lay the foundations for future BOSS studies, where the intrinsic distributions of 
spectroscopically-derived quantities can be determined. In the broader picture, an important 
goal of this paper is to complement the detailed characterization of the main statistical 
properties of the LRG population initiated in MD2014 that will be eventually 
used, in combination with N-body numerical simulations, to investigate the 
intrinsic clustering properties of these systems and the halo-galaxy 
connection in a fully consistent way. The connection between 
galaxies and halos will be performed by applying the techniques of halo occupation distributions 
(HOD: e.g., \citealt{Berlind2002}; \citealt{Zehavi2005}) and halo abundance matching (HAM: e.g., \citealt{Vale2004}; \citealt{Trujillo2011}).
These future applications are addressed in more detail in the last section of the paper.

\section{The data} 
\label{sec:data}

\subsection{The CMASS sample} 
\label{sec:cmass}

In this work we make use of both spectroscopic and photometric data from the Tenth Data Release of the SDSS 
(DR10, \citealt{Ahn2014}), which corresponds to the third data release of the SDSS-III program and the second 
release that includes BOSS data. We choose the DR10 instead of the recently 
published Data Release 12 (DR12, \citealt{Alam2015}), in order to be consistent with the luminosity 
function results shown in MD2014. The spectroscopic DR10 BOSS sample contains a total of $927,844$ galaxy 
spectra and $535,995$ quasar spectra (this is a growth of almost a factor two as compared to the SDSS DR9, \citealt{Ahn2012}). The baseline 
imaging sample is the final SDSS imaging data set, which contains, not only the new SDSS-III imaging, but also the previous SDSS-I and II imaging data.
This imagining data set was released as part of the DR8 \citep{Aihara2011}. These imagining programs provide five-band {\it{ugriz}} 
imaging over 7600 sq deg in the Northern Galactic Hemisphere and $\sim$ 3100 sq deg in the Southern Galactic 
Hemisphere. The typical magnitude corresponding to the $50\%$ completeness limit for detection of point sources is $r = 22.5$. The 
following papers provide comprehensive information about technical aspects of the SDSS survey:
\cite{Fukugita1996} describes the SDSS {\it{ugriz}} photometric system; 
\cite{Gunn1998} and \cite{Gunn2006} describe the SDSS camera and the SDSS telescope, respectively;
\cite{Smee2013} provides detailed information about the SDSS/BOSS spectrographs. 

The catalog that we used to compute the RS LF in MD2014 is the DR10 Large Scale Structure catalog (DR10 LSS). 
The DR10 LSS, which is basically built from the BOSS spectroscopic catalog and is thoroughly described in \cite{Anderson2014}, contains a small 
number of galaxies from the SDSS Legacy Survey. The SDSS Legacy Survey includes the SDSS-I survey and a small fraction of the SDSS-II survey. 
In the previous SDSS programs, the spectra were observed through 3 arcsec fibers, while the aperture size is only 2 arcsec for BOSS.
This varying aperture size makes the LSS catalog slightly heterogeneous as far as the velocity-dispersion measurement is concerned. 
To avoid this problem, we opt to use the spectroscopic catalog, which is very similar to the LSS catalog, and essentially maps the same 
intrinsic population as that described by the RS LF presented in MD2014. We have also checked that SDSS Legacy
galaxies are confined to a very small region of low redshift and high luminosity within the redshift-luminosity space. 
Importantly, they are mostly found at $z\lesssim0.5$, which is below the redshift range where our LF results are reliable ($0.52 \lesssim z \lesssim 0.65$, see MD2014).

 We restrict our analysis to the CMASS (for ``Constant MASS") spectroscopic sample of the BOSS spectroscopic catalog. This
sample is built within the official SDSS-III pipeline by first applying an imaging pre-selection to ensure that only detections passing a particular set of quality criteria are 
chosen as targets. Secondly, a set of colour-magnitude cuts is applied to the resulting catalog, 
intended to select the LRG sample required to effectively measure the BAO
within a nominal redshift range $0.43<z<0.70$ (the sample extends slightly beyond these limits). In a similar 
way, the low-redshift LOWZ sample (not used in this paper) is constructed, covering a nominal redshift range $0.15<z<0.43$.
For more information on the BOSS selection refer to \cite{Eisenstein2011}, \cite{Dawson2013} and \cite{Reid2015} (also a 
summary is provided in MD2014). Importantly, the CMASS selection allows for a fraction of $\sim37\%$ of blue cloud objects, as measured by 
MD2014. The stellar masses for the red population, as measured by \cite{Maraston2013},  
are $M_* \gtrsim 10^{11} M_{\odot}$, peaking at $M_* \simeq 10^{11.3} M_{\odot}$, assuming a Kroupa 
initial stellar mass function \citep{Kroupa2001}. All colours 
quoted in this paper are {\it{model}} colours and all magnitudes {\it{cmodel}} magnitudes. 

 The total number of unique CMASS galaxies with a good redshift estimate and with model and cmodel apparent magnitudes 
and photometric errors in all g,r and i bands  in the catalog is 549,005. The mean redshift in the sample is 
$0.532$ and the standard deviation $0.128$; approximately $\sim7.5 \%$ and $\sim4.5 \%$ of galaxies 
lie below and above the nominal low redshift and high redshift limits, i.e., $z=0.43$ and $z=0.70$, respectively.  
For a complete discussion on the selection effects affecting the CMASS data, see MD2014.

\subsection{Velocity dispersion likelihood functions}
\label{sec:vdisp}

One of the key ingredients in our L-$\sigma$ relation study is the likelihood function of the central stellar velocity dispersion. The method for determining 
this likelihood function is described in detail in \cite{Shu2012}. Here we provide a brief summary. For every galaxy in our
sample, the line-of-sight stellar velocity dispersion within the central circular region of radius $1$ arcsec, 
is measured spectroscopically by fitting a linear combination of broadened stellar {\it{eigenspectra}} to the observed galaxy 
spectrum (note that the typical seeing for BOSS is $1.5$ arcsec). Thus, for the $i^{th}$ galaxy, the $\chi^2$ of the fit as a function of the trial velocity dispersion is
converted into the likelihood function of velocity dispersion with respect to the observational data $d_i$ as
\begin{equation}
p (d_i | \log_{10} \sigma) \propto \exp[-\chi^2_i(\log_{10} \sigma)/2].
\end{equation}
The best-fit velocity dispersion and its uncertainty can then be inferred from the $\chi^2$ function. 

\begin{figure}
\centering
\includegraphics[width=8.5cm]{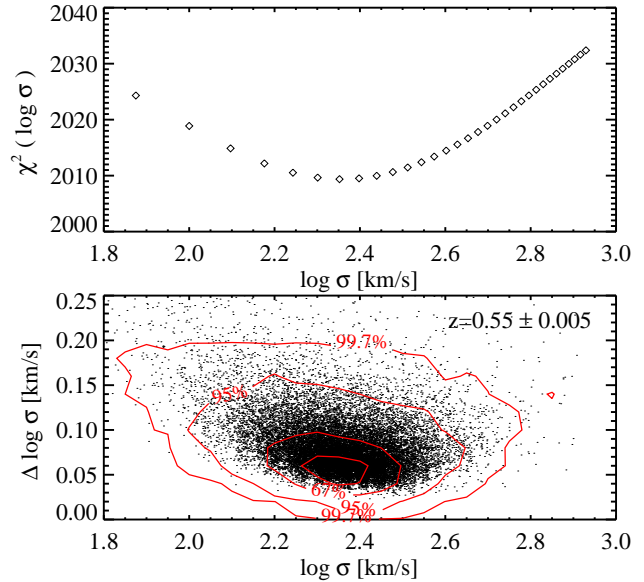}
\caption{
\label{fig:vdisp_likelihood}
An illustration of the BOSS velocity dispersion data. \textit{Top panel}: 
a typical $\chi^2 (\sigma)$ curve as a function of the trial $\sigma$ (D.O.F.=1736). 
This particular galaxy within the redshift range $z=0.55 \pm 0.005$ is chosen to 
have best-fit velocity dispersion and uncertainty matching the densest region shown in the 
bottom panel plot. \textit{Bottom panel}: scatter plot of the best-fit velocity dispersion in logarithmic scale 
and its uncertainty (also in logarithmic scale, black dots) for red CMASS galaxies ($g-i > 2.35$) within the redshift range $z=0.55 \pm 0.005$. 
Contours that enclose $30\%$, $67\%$, $95\%$, and $99.7\%$ of this subsample 
are overplotted in red.}
\end{figure}

Figure~\ref{fig:vdisp_likelihood} illustrates the type of velocity dispersion data 
that we have in BOSS. In the bottom panel, the best-fit central velocity dispersion 
and its uncertainty are shown in a scatter plot for all red CMASS galaxies 
within the redshift slice $z=0.55 \pm 0.005$. Here, a simple colour cut $g-i > 2.35$ is used
to approximately remove blue galaxies in observed space (following \citealt{Masters2011}). This is just illustrative,  
as in MD2014 we demonstrate that a simple colour cut is not efficient 
in terms of isolating the intrinsic RS, as large photometric errors scatter objects in and out of the  
colour demarcation. At $z=0.55$ the colour cut removes $20\%$ of the sample. The fraction 
of intrinsically non-RS objects in the CMASS sample at the same redshift according to the population 
deconvolution of MD2014 is $38\%$. 

The two quantities shown in Figure~\ref{fig:vdisp_likelihood} are inferred from the 
velocity-dispersion likelihood functions. Figure~\ref{fig:vdisp_likelihood} shows that the distribution is centered 
around $\log_{10} \sigma \simeq 2.35$, or 220 km/s, and $\Delta \log_{10} \sigma \simeq 0.06$ dex, or 30 km/s. More than two-thirds of 
the subsample have velocity dispersions between $120$ km/s and $300$ km/s 
with $20-50$ km/s uncertainties. The top panel displays an example of a typical $\chi^2$ function of a galaxy 
with both velocity dispersion and uncertainty consistent with the aforementioned central values. 

As the broad shape of the $\chi^2$ function displayed in the top panel of Figure~\ref{fig:vdisp_likelihood}
indicates (and as also emphasized in \citealt{Shu2012}), a point estimate for the velocity dispersion is 
only partially informative, even after a dedicated treatment is adopted for the new $\chi^2$ calculation. 
Note that an appreciable fraction of this subsample have best-fit velocity dispersions 
equal to $0$ km/s due to the limitations of fitting low SN spectra. Using the likelihood 
function propagates all information to the higher level analysis of the entire population. That is one of the main motivations for 
employing a hierarchical Bayesian statistical approach in this paper.

\section{Bimodality in the colour-colour plane: The intrinsic Red Sequence distribution} 
\label{sec:intrinsic}

 A key element in the computation of the intrinsic L-$\sigma$ relation for the RS from the CMASS sample is
the underlying intrinsic RS magnitude and colour distribution. This aspect is 
thoroughly addressed in MD2014, where we present an analytical method 
for deconvolving the observed (g-r) colour - (r-i) colour - (i)-band magnitude CMASS distributions 
from the blurring effect produced by photometric errors and selection cuts. The CMASS sample 
comprises a considerable fraction of bluer objects, that can scatter into the red side of the 
colour-colour plane due to photometric errors. In MD2014, this aspect is treated 
by modeling the BC and the RS separately (red/blue deconvolution), which allows us to correct the RS intrinsic distribution 
from the contamination caused by BC objects. Importantly, this modeling is not performed from the basis of 
previous assumptions about ``blue" or ``red" objects based on stellar population synthesis models,
but is intended to describe the bimodality found in the colour-colour plane. With this consideration, 
these components present the following characteristics:

\begin{itemize}
	\item {\it{Red Sequence (RS)}}: The RS is so narrow that is consistent, within the errors, at fixed magnitude and for a narrow redshift slice, with a delta function in 
	the colour-colour plane (width $<0.05$ mag), with only a shallow colour-magnitude relation shifting the location of this point. The results 
	reported in this paper regarding the L-$\sigma$ relation are based on this intrinsic distribution.
	\item {\it{Blue Cloud (BC)}}: The BC is defined as a background distribution that contains {\it{everything not
	belonging to the RS}} and is well described by a more extended 2-D Gaussian in the colour-colour plane. The
	RS is superimposed upon the BC, that extends through the red side of the colour-colour plane. Again, the name ``BC" is 
	not meant to imply that this distribution contains only blue, young objects; the BC is a spectroscopically 
	and photometrically heterogeneous population to which other types of ETGs can pertain, such as dusty ellipticals not belonging 
	to the narrow RS.  
\end{itemize}

The intrinsic distribution of magnitudes for the RS is the key ingredient in the computation of the RS L-$\sigma$ relation, and this is 
provided in MD2014 in the form of the RS LF. For the sake of convenience,  
and given the very narrow redshift slices that we consider, we will constrain the $\log_{10} \sigma - m_i$ (apparent 
magnitude in the i band) relation (note that this is not the L-$\sigma$ relation, which involves absolute magnitudes).
 As we show in the following section, for constraining this relation, it suffices to know the 
shape of the intrinsic distribution of apparent magnitudes, which is given by a Schechter Function of the form:

\begin{eqnarray} \displaystyle
n_{sch}(m_i,\phi_*,m_*, \alpha\}) = 0.4 log(10) \phi_* \left[10^{0.4(m_*-m_i)(\alpha+1)}\right] \times  \nonumber \\
exp\left(-10^{0.4(m_*-m_i)}\right) 
\label{eq:schechter}
\end{eqnarray}

\noindent where $\alpha = -1$ and $\phi_* =$ unity, for the sake of simplicity. The assumption 
that $\alpha = -1$ is dictated by the narrow magnitude range, that prohibits fitting for $\alpha$.
The intrinsic RS i-band magnitude distribution, as a function of 
redshift, can be obtained by inserting the following linear relation for the redshift evolution of the characteristic magnitude $m_*$ into 
Equation~\ref{eq:schechter}:

\begin{eqnarray} \displaystyle
m_*^{RS} (z) = (4.425\pm0.125)~(z-0.55) + (20.370\pm0.007)
\label{eq:linear_fits_rs}
\end{eqnarray}

\noindent Similarly, for the BC:

\begin{eqnarray} \displaystyle
m_*^{BC} (z) = (4.011\pm0.178)~(z-0.55) + (20.730\pm0.010)
\label{eq:linear_fits_bc}
\end{eqnarray}

 All the relations provided in this section come from the analysis presented in MD2014, 
although they are not explicitly reported there. 
By using these linear relations instead of the best fit values at each redshift we avoid introducing unnecessary noise into the analysis. 
These relations are obtained within the redshift range $0.525 < z < 0.65$, where selection 
effects are less severe. Note that the BC LF
was not reported in MD2014 due to the extreme incompleteness affecting the BC in the 
CMASS sample. Instead, it is used as a means of accounting for the contribution 
of the BC in the sample and of subtracting this contribution from the RS LF. Following the same strategy, we 
compute the $\log_{10} \sigma - m_i$ relation for the BC (and later the BC L-$\sigma$ relation) only as a means of subtracting the contribution of the
BC on the RS $\log_{10} \sigma - m_i$ relation. 

The last ingredient in the computation of the RS $\log_{10} \sigma - m_i$ relation, as we show in the next section, is the fraction of BC objects in the
CMASS sample, i.e., $f_{blue}$, which is also provided in MD2014. Again, to avoid 
introducing noise in the computation, we use the following linear relation for the redshift dependence of $f_{blue}$:

\begin{eqnarray} \displaystyle
f_{blue} (z) = (0.890\pm0.051)~(z-0.55) + (0.381\pm0.003)
\label{eq:linear_fits_bc2}
\end{eqnarray}

\section{PDSO: Formalism for the photometric deconvolution of spectroscopic observables} 
\label{sec:formalism0}

\subsection{The method}
\label{sec:formalism}

The L-$\sigma$ relation is a relation between a spectroscopic observable, 
the stellar velocity dispersion $\sigma$, and a photometric observable, the luminosity $L$ (or, 
in a more practical fashion, the $\log_{10} \sigma$ and the absolute magnitude $M$). As
previously mentioned, measuring velocity dispersions from BOSS galaxies is hindered by 
low SN spectra, which dictates that instead of point measurements we use likelihood
functions for the velocity dispersions. The photometric aspect of the L-$\sigma$ relation computation presents
challenges as well: as discussed in MD2014, the observed CMASS 
distribution is strongly affected by photometric errors and selection effects. In this section
we present a formalism that combines our likelihood measurements for the velocity dispersions and 
our red/blue deconvolution of photometric quantities. This formalism, which we call PDSO, can be used 
for the photometric deconvolution of any spectroscopic observables, so we will present it 
in a general way.

Let us start by assuming that we have a sample of galaxies indexed by $i$, each of which has a  spectroscopic data
vector $\mathbf{d}_i$ (in our case, velocity dispersions) and a photometric data vector $\mathbf{c}_i$ (colours and/or magnitudes). 
These galaxies are selected (i.e., for measuring the spectra)
according to some photometric cuts described by $P(\mathbf{c})$, which is
the probability (between 0 and 1) of selecting a galaxy with photometric
data vector $\mathbf{c}$. $P(\mathbf{c})$ in the case of the BOSS CMASS sample can be straightforwardly 
derived from the CMASS selection scheme (see \citealt{Dawson2013}). 

Assume that we have quantified the probabilistic mapping from \textit{intrinsic}
(noise-free) photometric values $\mathbf{C}$ into observed (noisy) photometric
values $\mathbf{c}$, which we denote as $p(\mathbf{c} | \mathbf{C})$. In MD2014
this mapping is obtained by modeling the covariance matrix 
of magnitudes and colours using SDSS Stripe 82 multi-epoch data. 

Assume that these galaxies have been quantified in terms of a number of
one or more photometrically distinct population components indexed by $k$,
with the results expressed as the intrinsic number density of galaxies
per unit $\textbf{C}$ in component $k$, denoted by $n_k(\mathbf{C})$
such that $n_k(\mathbf{C}) \, d\mathbf{C}$ has units of spatial number density. 
In reality, as we will show below, this quantity can also have units of number.
The above characterization can come in the form of the decomposition into RS and
BC galaxy colour--luminosity functions. Note that $n_k(\mathbf{C})$ need not be normalized over $\mathbf{C}$, and indeed may be divergent.

As a result of the previous analysis, we have derived the individual probability 
density functions (PDF's) of each component in observed space within the selected sample, as well
as the fraction of sampled objects in each component.  To express these in terms
of previously introduced quantities, we introduce the following function
for notational convenience:
\begin{equation}
\Phi_k(\mathbf{c}) = P(\mathbf{c}) \int d\mathbf{C} \, p(\mathbf{c} | \mathbf{C}) \, n_k(\mathbf{C}).
\end{equation}
This is the number (density) of galaxies in observed photometric space
for component $k$.

The number of galaxies in the sample in component $k$ is
\begin{equation}
N_k = \int d\mathbf{c} \, \Phi_k(\mathbf{c})
\end{equation}

The fraction of sample galaxies pertaining to component $k$ is
\begin{equation}
f_k = N_k \left/ \sum_{k\prime} N_{k\prime} \right. .
\end{equation}

Note that $f_k$ would correspond to $f_{blue}$ (and $1-f_{blue}$) within our deconvolution
framework. The PDF of galaxies in observed photometric space in component $k$ is
\begin{equation}
p_k(\mathbf{c}) = \Phi_k(\mathbf{c}) / N_k .
\end{equation}

The full PDF of the observed sample in observed photometric space is
\begin{equation}
p(\mathbf{c}) = \sum_k f_k \, p_k(\mathbf{c}).
\end{equation}

A derived function that we will make use of below is the PDF of a
sample galaxy in \textit{intrinsic} photometric space given its
position in \textit{observed} photometric space, which we can derive
by noting that the joint PDF of observed and intrinsic photometric vectors
for component $k$ is proportional as
\begin{equation}
p_k(\mathbf{C},\mathbf{c}) \propto P(\mathbf{c}) \, p(\mathbf{c} | \mathbf{C}) \, n_k(\mathbf{C})
\end{equation}

so that then
\begin{eqnarray}
p_k(\mathbf{C} | \mathbf{c}) &=& {{p_k(\mathbf{C},\mathbf{c})} \over {\int d\mathbf{C} \, p_k(\mathbf{C},\mathbf{c})}} \\
&=& {{P(\mathbf{c}) \, p(\mathbf{c} | \mathbf{C}) \, n_k(\mathbf{C})}
\over {\int d\mathbf{C} \, P(\mathbf{c}) \, p(\mathbf{c} | \mathbf{C}) \, n_k(\mathbf{C})}} \\
&=& {{p(\mathbf{c} | \mathbf{C}) \, n_k(\mathbf{C})}
\over {\int d\mathbf{C} \, p(\mathbf{c} | \mathbf{C}) \, n_k(\mathbf{C})}}
\end{eqnarray}

Importantly, $p_k(\mathbf{C} | \mathbf{c})$, which is the function that we will use ultimately, is independent of some 
of the choices that we make about the intrinsic distributions, $n_k(\mathbf{C})$. In particular, 
any normalization will cancel off, so we can simply use the intrinsic number counts shown in 
Equation~\ref{eq:schechter}, where volume is not taken into account and $\phi_* = 1$ unity
by definition. 

Now assume that there is some parameter or vector of parameters $\mathbf{x}$
that can be measured from the spectroscopic data vector $\mathbf{d}_i$ in the
sense that we can write down and compute the function
\begin{equation}
p(\mathbf{d}_i | \mathbf{x}_i)
\end{equation}
for each galaxy $i$.  In our case, $\mathbf{x}$ is the velocity dispersion,
$\sigma$ and $p(\mathbf{d}_i | \mathbf{x}_i)$ is proportional
to the function $\exp[-\chi^2(\sigma)]$.

Next we assume a parameterized model for the variation of the
spectroscopic observable within the sample populations as a function
of photometric values.  This is expressed as
\begin{equation}
p_k (\mathbf{x} | \mathbf{C} ; \mathbf{t}).
\end{equation}
The vector $\mathbf{t}$ will denote
the ``hyperparameters'' that describe this PDF.
Our goal is to infer the elements of $\mathbf{t}$.
To do this, we proceed to express the likelihood
function of $\mathbf{t}$ given the spectroscopic data
and the photometric data. In our framework,
the hyperparameters $\mathbf{t}$ only affect the
probabilities of the spectroscopic observables \textit{given}
the photometric observations.
\begin{eqnarray}
&&\!\!\!\!\! \mathcal{L} (\mathbf{t} | \left\{\mathbf{d}_i\right\}, \left\{\mathbf{c}_i\right\}) = p(\left\{\mathbf{d}_i\right\} | \left\{\mathbf{c}_i\right\}, \mathbf{t}) \\
&=& \prod_i p(\mathbf{d}_i | \mathbf{c}_i, \mathbf{t}) \\
&=& \prod_i \int d\mathbf{x} \, p(\mathbf{d}_i | \mathbf{x}) \, p(\mathbf{x} | \mathbf{c}_i, \mathbf{t}) \\
&=& \prod_i \int d\mathbf{x} \, p(\mathbf{d}_i | \mathbf{x}) \sum_k f_k \, p_k(\mathbf{x} | \mathbf{c}_i, \mathbf{t}) \\
&=& \prod_i \int d\mathbf{x} \, p(\mathbf{d}_i | \mathbf{x}) \sum_k f_k \int d\mathbf{C} \, p_k(\mathbf{x} | \mathbf{C}, \mathbf{t})
\, p_k(\mathbf{C} | \mathbf{c}_i)
\label{eq:likelihood}
\end{eqnarray}

At this point, we have arrived at an expression on the right-hand
side entirely in terms of quantities that we have introduced
above, and we can proceed to map and/or maximize the likelihood
function of the hyperparameters $\mathbf{t}$.

\subsection{Application to the computation of the L-$\sigma$ relation in BOSS} 
\label{sec:application}

 For the sake of convenience, we proceed by applying our formalism to the 
computation of the $\log_{10} \sigma -m_i$ relation from the CMASS sample. 
Our spectroscopic observable is therefore the logarithm of the 
velocity dispersion, i.e. $\log_{10} \sigma$, and the intrinsic distributions ($n_k$) are 
represented by the Schechter number counts of Equation~\ref{eq:schechter}, with a redshift 
dependence given by Equations~\ref{eq:linear_fits_rs} and \ref{eq:linear_fits_bc} for the
RS and the BC, respectively. This choice implies that $\mathbf{C}$ and $\mathbf{c}$ correspond 
to the i-band apparent magnitude, $m_i$, in intrinsic and observed space, respectively. 

The parameterized model for the variation of the spectroscopic observable ($\log_{10} \sigma$) 
within the sample populations as a function of photometric values ($m_i$) encodes the  
$\log_{10} \sigma-m_i$  relation. Motivated by results from \cite{Bernardi2003b}, we approximate the intrinsic distribution
of velocity dispersions at fixed L as a Gaussian distribution in $\log_{10} \sigma$ with mean $<\log_{10} \sigma>$ and intrinsic scatter $s$. For component k this has 
the form:

\begin{equation}
p_k (\log_{10} \sigma | \mathbf{m_i} ; \mathbf{t_k}) = \frac{1}{\sqrt{2 \pi} s_k} \exp\left[\frac{-(\log_{10} \sigma - <\log_{10} \sigma>_k)^2}{2 s_k^2}\right]
\label{eq:gaussian}
\end{equation}

The mean of the velocity dispersion for component k, i.e.,$ <\log_{10} \sigma>_k$, is assumed to follow 
a linear relation with apparent magnitude, $m_i$, of the form:

\begin{equation}
<\log_{10} \sigma>_k = c_{1,k} + 2.5 + c_{2,k} ( m_i - 19)
\label{eq:F-J relation}
\end{equation}

This expression, as we will show in following sections, can be easily transformed, within our framework, into 
the L-$\sigma$ relation, which is expressed in terms of absolute magnitudes.   

\section{Aperture Correction} 
\label{sec:aperture}

BOSS velocity dispersions are measured within the 2 arcsec diameter aperture of the BOSS fibers.  As we move to higher 
redshift within the CMASS sample, the angular size of the fiber probes progressively larger physical scales. This 
effect is accounted for {\it{a posteriori}}, by applying an aperture correction (AC) to the best-fit relations obtained by 
maximizing the likelihood function of Equation~\ref{eq:likelihood}. By assuming a de Vaucouleurs profile 
for the variation of the surface brightness as a function of apparent distance to the center of the galaxy, we have obtained 
the following relation:  

\begin{equation}
\sigma_{obs} / \sigma(<R_e) = 0.98~(R_e/R_{\mathrm{aperture}})^{0.048} 
\label{eq:aperture}
\end{equation}

\noindent that relates the observed velocity dispersion $\sigma_{obs}$ that we measure in BOSS, the velocity dispersion
averaged within the effective radius, $R_e$, and the effective radius itself, in arcsec. As part 
of the derivation of the above relation, the blurring produced by an average seeing of 1.8 arcsec has 
been assumed.

In order to perform a realistic AC we need to take into account the variation of $R_e$ as a function of apparent magnitude, $m_i$, for each redshift slice, 
which may affect the slope of the $\log_{10} \sigma -m_i$ relation. To this end, we fit a linear relation to the mean observed i-band $\log_{10} R_e$ measured by 
the pipeline (in arcsec) as a function of $m_i$ \footnote{\citealt{Beifiori2014} present
an analysis on the evolution of the effective radius in BOSS, but the $R_e$ - magnitude relation necessary 
to derive the aperture correction is not reported.}. For a given redshift slice, this relation takes the form:

\begin{equation}
<\log_{10} R_{e}> = a + b (m_i - 19)
\label{eq:r_e}
\end{equation}

\noindent where we have shifted the reference magnitude to $m_i = 19$, similarly to Equation~\ref{eq:F-J relation}, for consistency. 
Figure~\ref{fig:aperture_correction2} displays the observed i-band $\log_{10} R_e$
in arcsec as a function of $m_i$ for the redshift slice centered at $z=0.55$, in contours enclosing 
67$\%$, 95$\%$ and 99.7$\%$ of the entire sample, respectively. The squares show the mean 
values in magnitude bins of $0.1$ mag and the errors the 1-$\sigma$ scatter around the mean. The solid line 
shows the linear fit to the mean values that we use to correct, on average, our $\log_{10} \sigma -m_i$ relation.  

By using Equation~\ref{eq:aperture}, it can be easily demonstrated that the aperture-corrected $c_1$ parameter, i.e. $c_1^{ac}$, is related to parameter $a$
in Equation~\ref{eq:r_e} in the following way:

\begin{equation}
c_1^{ac} = c_1 - 0.048 a - \log_{10}(0.98)
\label{eq:m1_corrected}
\end{equation}

And, similarly, for $c_2^{ac}$:

\begin{equation}
c_2^{ac} = c_2 - 0.048 b
\label{eq:m2_corrected}
\end{equation}

Within the same redshift slice, more luminous galaxies are larger in size (see Figure~\ref{fig:aperture_correction2}), which 
implies $b<0$. The AC correction by definition tends, therefore, to steepen the L-$\sigma$ relation.

In order to avoid introducing any extra noise we use a linear fit to the values of $a$ and $b$ as a function 
redshift, $a(z)$, $b(z)$. This is shown in Figure~\ref{fig:aperture_correction1} for the redshift range 
of interest, $0.5<z<0.7$. The linear fit that we obtain for $a(z)$ is:

\begin{figure}
\begin{center}
\includegraphics[scale=0.43]{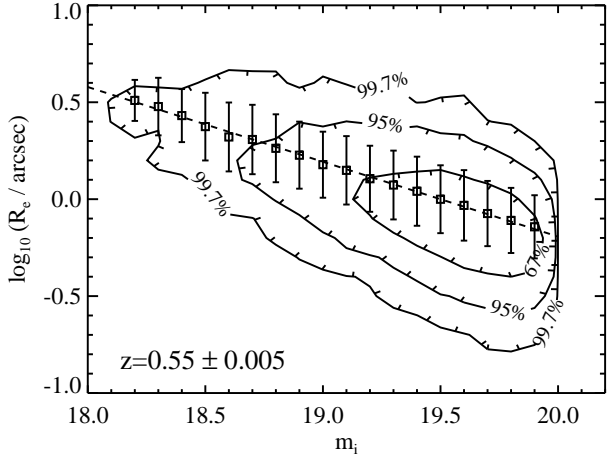}
\caption{The observed i-band $\log_{10} R_e$ in arcsec as a function of the i-band apparent magnitude
 $m_i$ for the redshift slice centered at $z=0.55$, in contours enclosing 
67$\%$, 95$\%$ and 99.7$\%$ of the entire subsample, respectively. The squares show the mean 
values in magnitude bins of $0.1$ mag and the errors the 1-$\sigma$ scatter around the mean. The solid line 
shows a linear fit to the mean values. }
\label{fig:aperture_correction2}
\end{center}
\end{figure} 

\begin{figure}
\begin{center}
\includegraphics[scale=0.43]{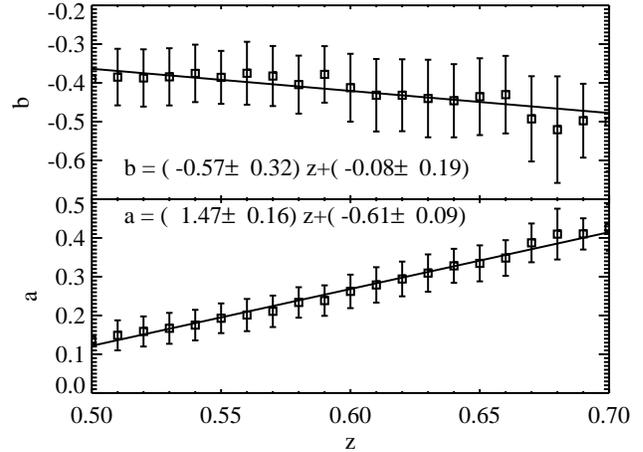}
\caption{The redshift dependence of parameters $b$ (upper panel, squares) and $a$ (lower panel, squares), representing, respectively, 
the slope and the zero-point of the mean relation between the observed i-band $\log_{10} R_e$ (in arcsec)
and the i-band apparent magnitude $m_i$. Errors correspond to uncertainty in the fit associated to each parameter.
The solid lines show the linear fits that we use to correct our L-$\sigma$ relation measurements.}
\label{fig:aperture_correction1}
\end{center}
\end{figure}

\begin{equation}
a(z) =  ( 1.47 \pm 0.16) z+( -0.61\pm 0.09)
\label{eq:a_z}
\end{equation}

\noindent and for $b(z)$:

\begin{equation}
b(z) = ( -0.57\pm 0.32) z+( -0.08\pm 0.19)
\label{eq:b_z}
\end{equation}

Note that the error on the slope of $b(z)$ is large, so we will consider also the case
where $b(z)$ is constant and equal to the mean value within the redshift range $0.52<z<0.65$, i.e. $b(z) =-0.402\pm0.020$. 

A typical value of $b(z) =-0.402$ implies a correction on the slope of the $\log_{10} \sigma$-$m_i$ relation (the $c_2$
parameter) of $\sim0.02$. Within our framework, the slope of the $\log_{10} \sigma$-$m_i$ relation at a given 
redshift slice coincides with the slope of the L-$\sigma$ relation. As a reference, typical values for this slope are $\sim-0.1$ (or equivalently 
4 for the exponent $\beta$ in the form $L \propto \sigma^\beta$; this is the F-J relation). This implies that the AC has a significant  
effect on the slope of the L-$\sigma$ relation when computed from BOSS.  The main idea is that brighter (and hence larger) 
galaxies have their velocity dispersions corrected by a different factor than fainter (and hence smaller)
galaxies, for a given fixed angular aperture, which affects the slope of the L-$\sigma$ relation.

 It is interesting to compare the effect of adopting a different aperture correction on the $\log_{10} \sigma -m_i$ relation. 
Equations~\ref{eq:m1_corrected} and \ref{eq:m2_corrected}, in combination with the values of parameters
$a(z)$ and $b(z)$, dictate the sensitivity of the AC to the exponent in Equation~\ref{eq:aperture}. By 
varying this exponent between a reasonable range, we can evaluate the effect on the 
zero-point and slope of the  $\log_{10} \sigma -m_i$ relation. We have chosen a range between $0.04$ (corresponding 
to the value derived by \citealt{Jorgensen1995}) and $0.066$ (found by \citealt{Cappellari2006}). The majority 
of values adopted in the literature are within this range (e.g. \citealt{Mehlert2003} estimate a value of $0.06$). 

 Using Equations~\ref{eq:m1_corrected}, and given that the typical variation 
of $a(z)$ within the redshift range considered is $\sim0.2$ (see Figure~\ref{fig:aperture_correction1}), we find that the 
variation in the AC for the zero-point, $\Delta AC$, within the redshift range considered (note that $a$ is a function of redshift),
for an exponent of $0.048$, is $0.0096$ dex, in absolute values. Increasing the exponent of the AC function to $0.066$ would translate into an increase in 
$\Delta AC$, for the zero-point, of a $38\%$. On the other hand, decreasing the exponent of the AC function to $0.04$ would result 
in a $\Delta AC$ $17\%$ smaller. Although the net effect is small, these variations can modify the zero-point - redshift trend.
In Section~\ref{sec:F-J relation} we discuss the redshift evolution of the zero-point, concluding that the effect on 
the zero-point of the uncertainties on the AC is significant, given the narrow redshift range that we probe and the 
mild evolution that we measure for the zero-point.

 With regard to the slope, from Equation~\ref{eq:m2_corrected} and Figure~\ref{fig:aperture_correction1}, as mentioned 
above, we find that the typical AC on this parameter is $\sim 0.02$ (i.e. $b(z) \times 0.048$), basically independent of redshift. Adopting a 
range of values for the exponent of the AC correction $0.04-0.066$ would translate into corrections within the range $0.016-0.026$. Although the 
effect is not negligible, adopting a different AC would not modify the main conclusion of the paper, in terms of the steep slope of the L-$\sigma$ 
relation, in any qualitative way (see following sections).

\section{Results} 
\label{sec:results}

\subsection{Best-fit parameters for the $\log_{10} \sigma$-$m_i$ relation} 
\label{sec:parameters}

By maximizing the likelihood function of Equation~\ref{eq:likelihood}, we obtain
the best-fit values for the hyperparameters $\mathbf{t}$. These parameters are the following, for each component $k$ (RS and BC): 
$c_{1,k}$, the zero-point of the $\log_{10} \sigma$-$m_i$ relation, that corresponds to the mean $\log_{10} \sigma$ 
at $m_i = 19$; $c_{2,k}$, the slope of this linear relation, and $s_k$, the intrinsic scatter. The optimization of the likelihood function has been performed 
from $z=0.40$ to $z=0.70$, using a bin size of $\Delta z = 0.01$. This redshift
range exceeds the redshift interval where the computation of the RS LF is more reliable, i.e. $0.52 \lesssim z \lesssim 0.65$.  
Results outside this high-confidence range are obtained by extrapolation of the linear fits to the redshift evolution of the intrinsic distribution 
and the fraction of blue objects in the sample.  

\begin{figure*}
\begin{center}
\includegraphics[scale=0.5]{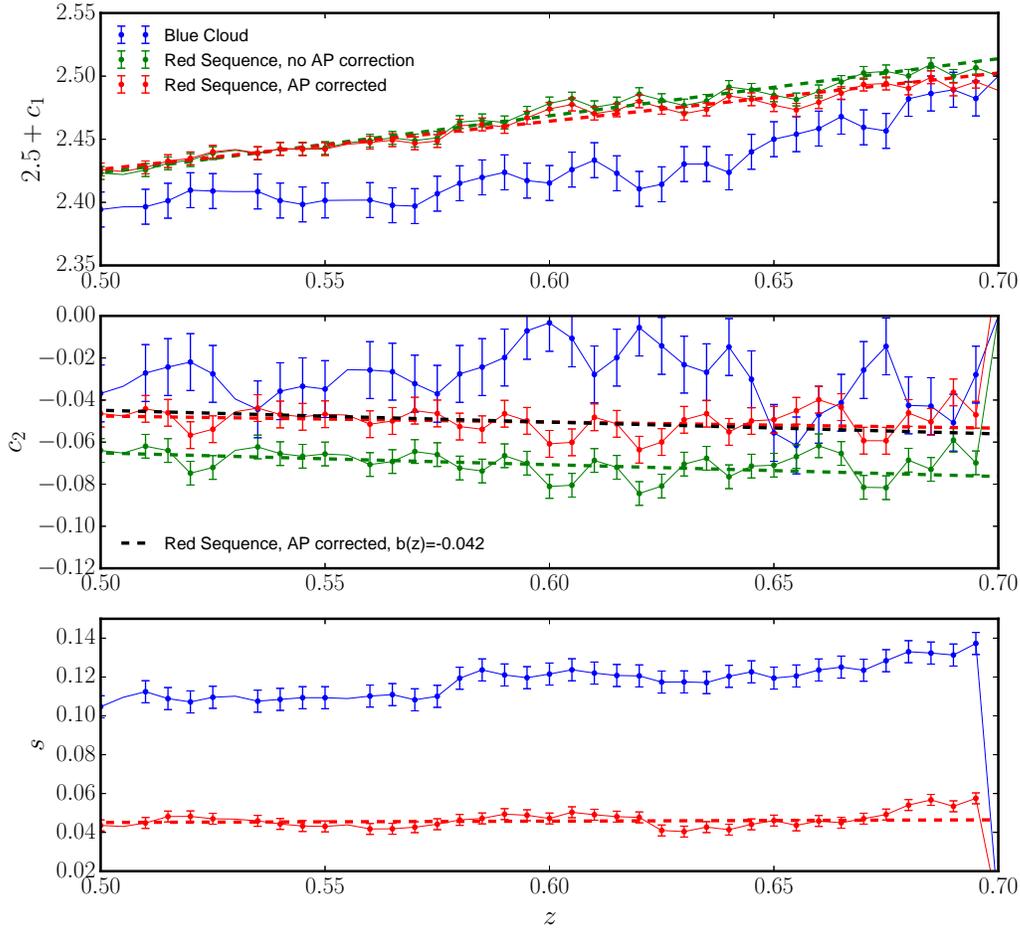}
\caption{The best-fit values for the hyperparameters $\mathbf{t}$ that describe the $\log_{10} \sigma$-$m_i$ relation
for both the RS and the BC component as a function of redshift. The distribution of velocity dispersions, for each component, 
is approximated as a log-normal distribution in $\log_{10} \sigma$ with mean $<\log_{10} \sigma>$ and intrinsic scatter $s$. 
The mean is parametrized as $<\log_{10} \sigma> = c_{1} + 2.5 + c_{2}~( m_i - 19)$.
In each panel, dots and solid lines represent the best-fit values obtained by maximizing the likelihood function of Equation~\ref{eq:likelihood}, 
following our hierarchical Bayesian analysis. Blue lines/symbols represent the $\log_{10} \sigma$-$m_i$ relation for the BC component, 
red lines/symbols show this relation for the RS component, once aperture correction is performed, and 
green lines/symbols show our results before aperture correction. Dashed lines show the best-fit linear models obtained within the redshift range considered.
The black dashed line in the second panel represents a linear fit where aperture correction to $c_{2}$ is assumed to be independent of redshift and 
equal to $b(z) = -0.042$. Errors on the best-fit parameters are assumed to be equal to the standard deviation with respect to the
best-fit model in each parameter.}
\label{fig:best_fit}
\end{center}
\end{figure*}

Figure~\ref{fig:best_fit} displays the redshift evolution of the 3 best-fit hyperparameters $\mathbf{t}$, 
for both the RS and the BC. It is important to bear in mind
that, due to extreme incompleteness, our BC component is not representative 
of the entire BC population, so we simply use it as a means of correcting for the
contribution of BC objects in the sample. In each panel, this BC component is represented in  
blue lines/symbols. The aperture-corrected RS parameters are represented in red and
the uncorrected RS parameters, in green.

Statistical errors on each parameter are obtained by mapping the likelihood function within the 6-dimension hyperspace 
and computing the posterior probability distribution for each parameter. This computation yields very small errors, which is 
consistent with the rapid convergence of the algorithm. While this procedure might not take all
the possible sources of error into account, other more realistic options, such as a bootstrap analysis, are 
computationally challenging. In order to avoid underestimating our uncertainties, we take a conservative approach here and, instead of the aforementioned 
error, we use, as the real error, the scatter on each parameter (standard deviation) with 
respect to the best-fit relation with redshift.

In the upper panel of Figure~\ref{fig:best_fit}, the 2.5 + $c_{1, RS}$ increases linearly with redshift, from a value 
of $\sim$ 2.43 at $z=0.50$ to $\sim$ 2.53 at $z=0.70$. The aperture correction has very little effect on the 
zero-point of the $\log_{10} \sigma$-$m_i$ relation. The following linear relation is obtained for the 
aperture-corrected RS $c_{1}$ (hereafter we drop the $ac$ superindex for simplicity): 

\begin{equation}
2.5 + c_{1,RS} (z) = (2.235\pm0.009) + (0.381\pm0.016)~z
\label{eq:m1}
\end{equation}

At fixed apparent magnitude ($m_i = 19$), we look at progressively more luminous galaxies as we 
move to higher redshift. The L-$\sigma$ relation implies that more luminous RS galaxies have higher velocity dispersions, which 
explains the $c_{1,RS}$ - redshift trend. On the other hand, the BC represents a photometrically and spectroscopically 
heterogenous population that contains a large fraction of blue, spiral galaxies (or even disky ellipticals) for which we expect 
smaller velocity dispersions. At fixed absolute magnitude, the BC component is expected, therefore, to have smaller mean $\log_{10} \sigma$
than the RS, which is exactly what the upper panel of Figure~\ref{fig:best_fit} shows. 

The middle panel of Figure~\ref{fig:best_fit} shows the redshift evolution of the slope of the $\log_{10} \sigma$-$m_i$ relation, $c_{2}$. The RS
is consistent with a single point in the colour-colour plane, with a shallow colour-magnitude relation that we neglect in the computation of
K-corrections. Absolute magnitudes are simply obtained by rescaling the apparent magnitudes at each redshift slice. This result 
implies that $c_{2,RS}$ coincides with the slope of the L-$\sigma$ relation, as we explicitly show in the next section. Our results for $c_{2,RS}$
indicate little evolution in the slope of the $\log_{10} \sigma$-$m_i$ within the redshift range considered. Importantly, 
the AC, as discussed previously, has a significant effect on $c_{2,RS}$, i.e., that 
changes from $\sim-0.07$ to $\sim-0.05$. The following linear relation with redshift is obtained for 
the aperture corrected $c_{2,RS}$:

\begin{equation}
c_{2,RS} (z) = - (0.033\pm0.012) - (0.029\pm0.021)~z
\label{eq:m2}
\end{equation}

 As mentioned above, this dependence of $c_{2,RS}$ with redshift is not significant given the scatter in the data. 
The $\Delta \chi^2$ with respect to the redshift-independent assumption is only $1.57$ (i.e. a significance 
of $1.25 \sigma$). The slope in the $\log_{10} \sigma - m_i$ relation that we measure within the redshift range $0.5<z<0.7$ is therefore 
$-0.070\pm0.006$ before AC and $-0.050\pm0.007$ after. The middle panel of Figure~\ref{fig:best_fit} also shows in a black dashed line our results 
for $c_{2,RS}$ assuming a constant value for the AC parameter $b$ of $b(z) = -0.042$. Neglecting the redshift 
evolution of $b$ in Figure~\ref{fig:aperture_correction1} has little impact on $c_{2,RS}$.

The bottom panel of Figure~\ref{fig:best_fit} displays the redshift evolution of the intrinsic scatter in the $\log_{10} \sigma - m_i$ relation for 
both the RS and the BC component.  This is the intrinsic RMS scatter in $\log_{10} \sigma$, at fixed $m_i$. Note that
an ``orthogonal" version of our result (i.e., scatter perpendicular to the $\log_{10} \sigma - m_i$ relation) would be essentially the same, given the steep slope
that we measure for this relation. The scatter for the RS, $s_{RS}$, increases slightly with redshift, but this trend is not
significant given the computed errors (the $\Delta \chi^2$ with respect to the redshift-independent assumption is only $1.36$, i.e., a significance 
of $1.17 \sigma$). The mean value that we obtain is only $s_{RS} = 0.047\pm0.004$ in $\log_{10} \sigma$.  This value
for the scatter in the $\log_{10} \sigma - m_i$ relation is very small as compared to the typical value of $\sim0.1$ found 
at intermediate-mass ranges and low redshift. In Section~\ref{sec:discussion} we show his this result is in excellent 
agreement with previous low-z high-mass results.

The hyperparameter $s$ for the BC component is significantly larger, i.e., $\sim 0.12$. Keep in mind that the BC component in our sample 
is by no means complete, and therefore, the corresponding $s$ value that we measure does not represent the true 
scatter for the entire BC population. It does provide, however, some indication of the scatter of the BC population 
relative to the RS population. The value of $\sim 0.12$ is consistent
with the intrinsic scatter trends shown by \citet{Shu2012}, who without accounting for the 
contamination effect caused by the BC objects, find that the overall intrinsic scatter value for 
the CMASS galaxy sample is $\sim 0.1$, an intermediate value due to the mixture of both RS and BC
galaxies. A trend of higher intrinsic scatter at higher redshift is also discovered by \citet{Shu2012}, which 
can be explained by the fact that the relative fraction of the BC objects in the CMASS sample increases 
with redshift. 

\begin{figure*}
\begin{center}
\includegraphics[scale=0.8]{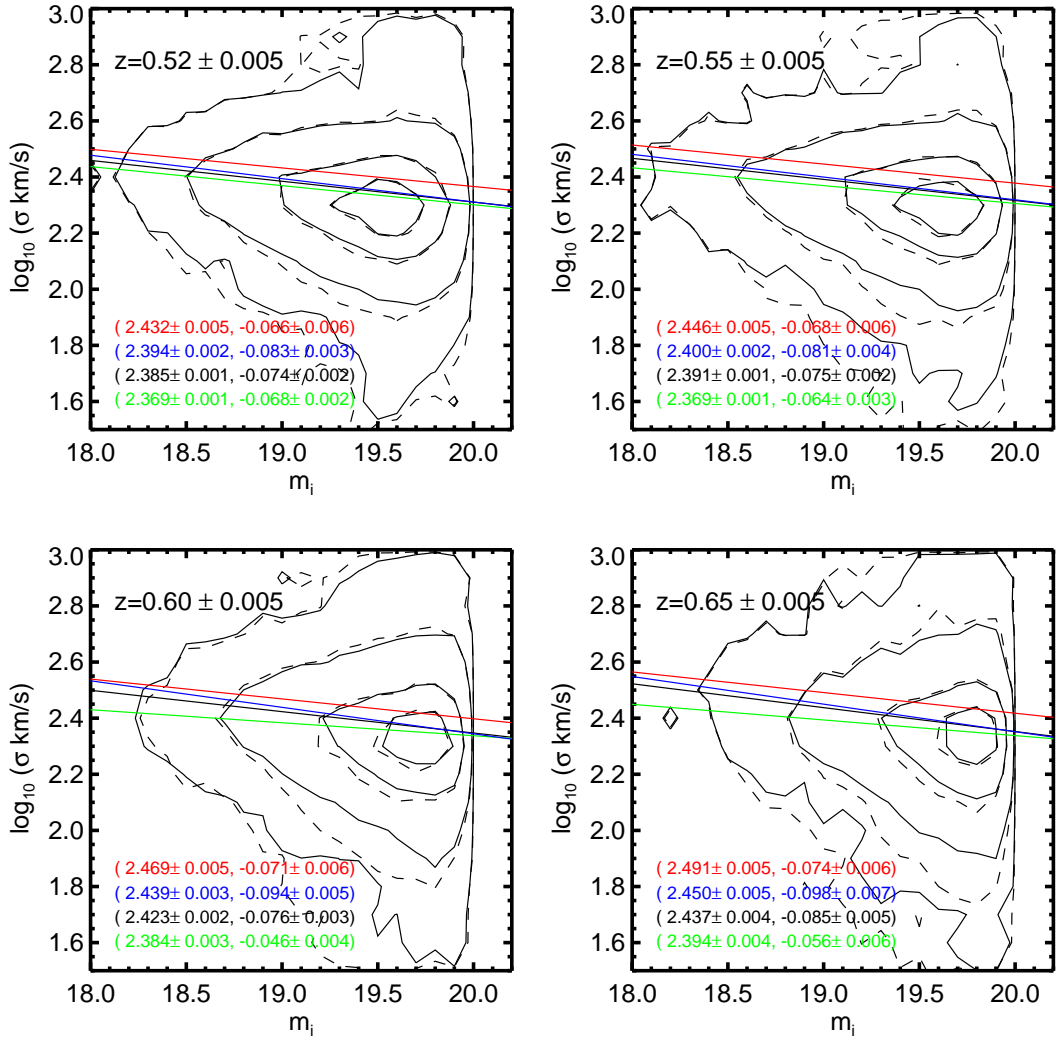}
\caption{The $\log_{10} \sigma - m_i$ relation as obtained using 4 different
methods in 4 different redshift slices ($z=0.52, 0.55, 0.60, 0.65$).  Results from Method 1, 2, 3, and 
4 (see text) are shown in red, blue, black and green, respectively. The solid contours show the best-fit central velocity dispersions 
(point estimates) as a function of apparent magnitude once a colour cut $g-i > 2.35$ is applied to remove blue objects, 
and the dashed contours represent the entire sample. The best-fit values for the slope and zero-point of the $\log_{10} \sigma - m_i$ 
are also provided.}
\label{fig:effect1}
\end{center}
\end{figure*}

 The $\log_{10} \sigma$-$m_i$ relation in a magnitude-$\log_{10} \sigma$ diagram for 4 different redshift 
slices is explicitly shown in Figure~\ref{fig:effect1} (Method 1, see next section).

\subsection{The effect of the various corrections implemented} 
\label{sec:effect}

Our computation of the $\log_{10} \sigma - m_i$ (equivalent to the L-$\sigma$ relation) at $z\sim0.55$ incorporates accurate treatments of
various issues affecting the BOSS data. In particular, we correct for 
incompleteness within the colour-colour-magnitude space, and we separately 
model the intrinsic RS and BC populations, which allows the 
statistical removal of all BC in the CMASS sample, including those that are 
predicted to scatter through the red side of the colour-colour plane (see MD2014
for more details). This information is incorporated into our PDSO method, a hierarchical Bayesian statistical 
framework that allows us to utilize the stellar velocity dispersion likelihood functions of 
individual objects, instead of the point estimates associated to them. This section is intended to 
provide a sense as to what impact these corrections have on our $\log_{10} \sigma - m_i$ results.

 In order to illustrate the effect of each correction, we have computed the $\log_{10} \sigma - m_i$ relation using the following 4 methods:

\begin{itemize}
	\item {\bf{Method 1}}: This is our optimized method, including all the corrections mentioned above.
	\item {\bf{Method 2}}: This uses exactly the same methodology as the one implemented in \cite{Shu2012}, applied to the dataset used in this work. \cite{Shu2012} developed of a hierarchical approach that incorporates velocity dispersion likelihood functions. The main differences with Method 1 are that 1) the whole 
CMASS sample is used for the computation, without any red-blue deconvolution, and 2) only a very approximative 
completeness correction is applied. For the sake of comparison, here we exclude blue
objects by imposing a simple colour cut $g-i > 2.35$ in observed spaced.
	\item {\bf{Method 3}}: The observed L-$\sigma$ relation for the blue subsample defined using a simple colour cut $g-i > 2.35$. No red-blue 
	deconvolution or completeness correction is applied. Instead of velocity dispersion likelihood functions, point estimates for the velocity dispersion are used. 
	\item {\bf{Method 4}}: Same as Method 3, but for the entire CMASS sample. 
\end{itemize}

\subsubsection{Effect on the slope and the zero-point}

\begin{figure}
\begin{center}
\includegraphics[scale=0.4]{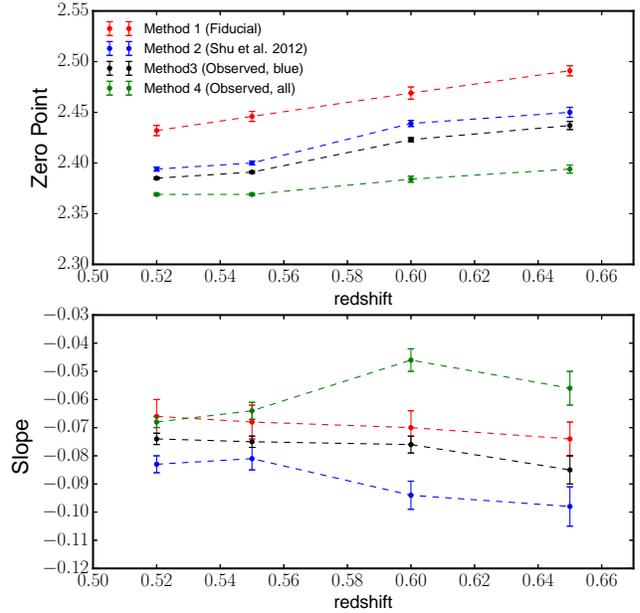}
\caption{ Zero-point (upper panel) and slope (lower panel) as a function of redshift obtained using the 4 different 
methods explained in the text (results from Method 1, 2, 3, and 4 are shown in red, blue, black and green, respectively).}
\label{fig:effect2}
\end{center}
\end{figure}

 In Figure~\ref{fig:effect1} and Figure~\ref{fig:effect2} we compare in 4 different redshift slices ($z=0.52, 0.55, 0.60, 0.65$) 
the $\log_{10} \sigma$-$m_i$ relation computed using the 4 methods presented above. Figure~\ref{fig:effect1} displays 
these relations in a $\log_{10} \sigma$ vs. $m_i$ diagram, where the solid contours show the best-fit central velocity dispersions 
(point estimates) as a function of apparent magnitude once the colour demarcation is applied, and the dashed contours represent the entire sample. 
Figure~\ref{fig:effect2} shows the redshift trends for the zero-point and the slope (not aperture-corrected, for simplicity), respectively. 

 With regard to the zero-point, Figure~\ref{fig:effect1} and especially the upper panel of Figure~\ref{fig:effect2}, indicate that 
an inadequate or partial removal of blue objects in the sample tends
to artificially push this parameter towards smaller values. This is expected, given that the velocity dispersion 
is obviously smaller in bluer (typically spiral) galaxies (see also the upper panel of Figure~\ref{fig:best_fit}). A partial 
removal of blue objects using a colour cut (Method 2) only palliates this effect slightly. We would still measure a zero 
point $10\%$ smaller than that of the intrinsic RS distribution. Part of this difference can also be due to the fact that in
Method 2 completeness is also only partially addressed, by just applying a rough correction to account for the scatter of objects in and out different 
magnitude bins, due to photometric errors. Interestingly, a comparison between the zero-point
obtained from Method 2 and Method 3 shows that, for the same sample (both using a colour cut in observed space, no deconvolution), 
the use of velocity dispersion likelihood functions in the context of a hierarchical Bayesian approach (Method 2) has a minor 
effect on the measured zero-point.

 The effect of corrections on the pre-aperture-corrected slope of the $\log_{10} \sigma - m_i$ relation is less obvious. The 
lower panel of Figure~\ref{fig:effect2} shows that the measured slope obtained with the optimized Method 1 is only slightly 
steeper (larger in this figure) than what we would measure using the observed distribution alone (with a colour cut, i.e., Method 3). This
differences would likely be within the errors once the slope is aperture corrected (see following section). Interestingly, using 
\cite{Shu2012} method, i.e. Method 2, would lead to a pre-aperture-corrected slope much closer to the canonical value of 4 (i.e. $-0.1$ in this
figure), $15 - 25 \%$ shallower than the values obtained with Method 1 (note that we would still measure a slope steeper than the canonical value once aperture 
correction is applied).

\subsubsection{Effect on the scatter}

The use of velocity dispersion likelihood functions in combination with completeness/intrinsic 
distribution results within a hierarchical Bayesian statistical framework has a tremendous impact on 
our ability to recover the intrinsic scatter in the $\log_{10} \sigma - m_i$ relation (which coincides 
with the scatter in the L-$\sigma$ relation). The typical observed scatter in Figure~\ref{fig:effect1} ranges from 
$\sim0.1$ at the bright end to $\sim0.16$ at the faint end. For the entire, partially-completeness-corrected 
observed distribution, \cite{Shu2012} report a value of $\sim0.1$, which is in agreement with the scatter 
in the higher-SN regime (within the sample). Our comprehensive analysis 
allows us to dig even deeper, unveiling the intrinsic scatter of the intrinsic RS distribution: a tiny $0.05$ dex,
as Figure~\ref{fig:best_fit} shows. 

 The difference between the scatter that we measure and the one reported by \cite{Shu2012} illustrates 
the importance of the red-blue deconvolution. As shown in the bottom panel of Figure~\ref{fig:best_fit}, the 
scatter that we expect for the BC population is significantly larger. This is consistent with the fact that the BC is 
a much more extended distribution (photometrically and spectroscopically heterogenous) in the colour-colour plane. Mixing RS and BC objects and not properly 
correcting for completeness inevitably leads to an increase in the reported intrinsic scatter of the L-$\sigma$ relation.

 In order to demonstrate that we actually have the statistical power to measure such 
a small scatter, we have performed a simple Monte Carlo variance-estimation analysis to estimate the resolution that 
we can expect given the number of objects that we have in a typical redshift bin ($\sim20,000$ objects)
and the typical error that we have in our individual velocity dispersion measurements ($\sim0.1$ dex, as Figure~\ref{fig:vdisp_likelihood}
shows). By simulating the observed distribution of velocity dispersions assuming an intrinsic scatter of $0.047$ dex in $\log_{10} \sigma$ and the
aforementioned typical measurement error on $\log_{10}\sigma$, we can evaluate our capacity to recover the intrinsic scatter. 
Our analysis shows that the typical uncertainty on the intrinsic scatter is of the order 
of $0.001$ dex, which is more than one order of magnitude smaller than the value that we find 
for the scatter, i.e. $0.047$ dex in $\log_{10} \sigma$. Note that the error that we estimate for the scatter is $0.004$ dex. This test gives us 
confidence that our measurement is not an artifact that results from working below our resolution limit.

\subsection{The L-$\sigma$ relation} 
\label{sec:F-J relation}

Translating the best-fit hyperparameters shown in Figure~\ref{fig:best_fit} into the standard L-$\sigma$ relation 
($\log_{10} \sigma$ as a function of absolute magnitude) for the RS is straightforward, due to the intrinsic characteristics of the 
RS colour-colour-magnitude distribution. In MD2014 we show that, at a given narrow redshift slice
(width $\Delta z = 0.01$), and magnitude bin, the RS intrinsic distribution is consistent with a
single point in the colour-colour plane, with only a shallow colour-magnitude relation that shifts this point slightly with L. 
Under such conditions, the K-correction, independently of the stellar population synthesis model chosen, changes very little within 
the apparent magnitude range of the CMASS sample, so we can basically assume it to be constant. We can, therefore, convert 
from apparent magnitudes to absolute magnitudes (K-corrected to $z=0.55$) at a given redshift slice by simply rescaling the 
apparent magnitude using the standard equation:

\begin{equation}
^{0.55}M_i = m_i - DM(z) - ^{0.55}K_i (z)
\label{eq:absmag}
\end{equation}

The width of each redshift slice is small enough that the variation of DM within the redshift bin can 
also be neglected. By substituting Equation~\ref{eq:absmag} into Equation~\ref{eq:F-J relation}, rearranging 
and adding a factor $M_0 c_{2}$ at each side of the equation we arrive at the following expression:

\begin{equation}
<\log_{10} \sigma> = c_{1}^{\prime} + 2.5 + c_{2}^{\prime} \left(^{0.55}M_i - M_0\right)
\label{eq:F-J relation2}
\end{equation}

\noindent where:

\begin{eqnarray} \displaystyle
c_{1}^{\prime} (z) =  c_{1} (z) -  c_{2} (z) (19 - DM(z) - ^{0.55}K_i (z) - M_0) \nonumber  \\
c_{2}^{\prime} (z) =  c_{2} (z)
\label{eq:F-J relation3}
\end{eqnarray}

The slope of the  L-$\sigma$ relation is independent of the fact that we use apparent magnitudes or absolute magnitudes, 
again due to the characteristic shape of the RS in colour-colour-magnitude space. The K-correction as a function 
of redshift, $^{0.55}K_i (z)$, is computed using a grid of models generated using the Flexible Stellar Population Synthesis  code (FSPS, \citealt{Conroy2009}) in the way
described in MD2014. This grid expands a range of plausible stellar population properties for redshift-dependent models 
within the CMASS redshift range. An average K-correction assuming the colours of the RS is computed at every redshift. 

\begin{figure}
\begin{center}
\includegraphics[scale=0.3]{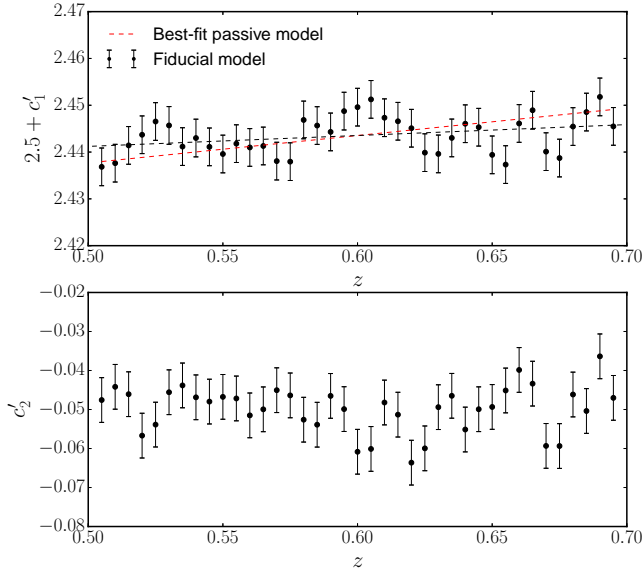}
\caption{ The aperture-corrected L-$\sigma$ relation parameters as a function of redshift within the redshift range $0.5<z<0.7$. Upper panel:
The zero-point or, equivalently, the mean of the $\log_{10} \sigma$ at a K-corrected reference magnitude of $^{0.55}M_i = -23$ (black dots),
and the best-fit linear model its redshift evolution (black dashed line). In addition, we show the best-fit passive model 
that fades at a rate of $1.18$ mag per unit redshift (as measured in MD2014). Lower panel: the slope
of the L-$\sigma$ relation, $c_{2}^{\prime}$. Errors are assumed to be equal to the standard deviation with respect to the 
best-fit model for each parameter. }
\label{fig:F-J relation_absmag}
\end{center}
\end{figure}

Figure~\ref{fig:F-J relation_absmag} displays parameters $c_{1}^{\prime}$ and $c_{2}^{\prime}$ for the  L-$\sigma$ relation
as a function of redshift, assuming $^{0.55}M_0 = -23$ (black). Errors are assumed to be equal to the standard deviation 
of the data for each parameter. The value of $^{0.55}M_0 = -23$ for the reference absolute magnitude has been 
chosen because it falls within the CMASS magnitude range across the entire redshift range considered (see MD2014). 

 As Figure~\ref{fig:F-J relation_absmag} indicates, the zero-point that we measure for the L-$\sigma$ relation 
has a slight dependence on redshift, so that larger values are found at higher redshifts. It is, however, 
a very small effect, of $\sim0.005$ dex within the redshift range considered ($0.5<z<0.7$, $\sim 1.3$ Gyr of cosmic time). The best-fit linear relation that we
measure is $2.429\pm0.007 + (0.023\pm0.011)~z$ (black dashed line). This redshift dependence is significant as 
compared to a best-fit redshift-independent value of $c_{1}^{\prime} = 2.443\pm0.004$ if we consider the whole redshift range ($\Delta \chi^2 = 4.49$, i.e., a significance 
of $\sim2 \sigma$), but not if we restrict ourselves to the high-confidence redshift 
range $0.52<z<0.65$ ($\Delta \chi^2 = 1.579$, i.e., a significance of $\sim1.25 \sigma$).

 In MD2014 we conclude that the LF evolution of the LRG population at $z\sim0.55$ is consistent with 
that of a passively-evolving population that fades at a rate of $1.18$ mag per unit redshift. Assuming plausible single-stellar population 
models, including both Flexible Stellar Population Synthesis (FSPS, \citealt{Conroy2009}) and M09 \citep{Maraston2009} models, such 
fading rate, at that redshift, translates into a formation redshift for the LRGs of $z=2-3$. The evolution of 
the zero-point assuming a best-fit passive model of the aforementioned characteristics is shown 
in a red dashed line in Figure~\ref{fig:F-J relation_absmag}. Here, by best-fit model we mean that the normalization 
of the zero-point of the passive model is fit to the data points, so only the redshift evolution is meaningful. 
The deviations found between our fiducial model (the best-fit
model that we obtained from our photometric deconvolution formalism) and the best-fit passive model
are of the order of $\pm0.004$ dex within the entire redshift range $0.5<z<0.7$. These deviations are again 
significant ($\Delta \chi^2 = 10.24$, i.e., a significance of $\sim3.2 \sigma$) if we considered the entire redshift range, but that 
significance is questionable if we restrict ourselves to the high-confidence redshift range ($\Delta \chi^2 = 3.01$, i.e., a significance of $\sim1.73 \sigma$).

The fact that the significance of the discrepancies with the best-fit passive model reported in MD2014 
depend strongly on the redshift range considered, in combination with the fact the effect is so small ($\pm 0.004$ dex),
appear to suggest that our results for the zero-point - redshift trend are consistent with the best-fit passive model. This 
idea is reinforced when we consider the uncertainty on the computation of the AC. A modification of the AC, within reasonable limits, would produce an effect 
of the same magnitude as the discrepancies that we find between the data and the best-fit passive 
model. As a matter of fact, a variation of the $a(z)$ and $b(z)$ functions involved in the AC, within the 
reported uncertainties, could account for the discrepancies found. More importantly, the trend shown in 
Figure~\ref{fig:F-J relation_absmag} is sensitive to the exponent of the AC, as discussed in Section~\ref{sec:aperture}. 
We have checked that by increasing slightly this exponent we can rapidly reach a much better 
agreement between the model and the data. A value of $0.06$ (instead of $0.048$), which is within the typical 
range of values previously used in literature, would produce almost a perfect agreement 
(a tiny $\Delta \chi^2 = 0.11$, when using the entire redshift range).

In summary, we conclude that our results are consistent with a passive model with formation 
redshift $z=2-3$, given the small variation of the zero-point that we measure within such a narrow redshift 
range and the uncertainties that the AC is subject to. This result is 
in good agreement with a variety of studies on the evolution of the FP, as we discuss in Section~\ref{sec:comparison}. 
An improvement in the AC or $R_e$ estimation in BOSS will be necessary to further 
constrain the evolution of the zero-point. The possibility remains that the discrepancies found may  
also be partially due to the fading rate of the passive model being too high; a value closer 
to $\sim0.9$ mag per unit redshift would suffice to make these discrepancies clearly not significant. This, 
in any case, would not imply a formation redshift much higher than $z=2-3$, given the typical fading-rate 
evolution of passive models (see MD2014 for a discussion).

\section{Discussion} 
\label{sec:discussion}

\subsection{Comparison with previous results} 
\label{sec:comparison}

In this section we show that our results for the high-mass L-$\sigma$ relation at $z\sim0.55$ are in 
good agreement with previous findings at low redshift.

The canonical form of the L-$\sigma$ relation for our chosen bandpass is $L_i \propto \sigma^{\beta}$,
where $L_i$ represents the i-band luminosity. The slope of this relation, $\beta$, is directly related to the measured slope
of the $\log_{10} \sigma$-$M_i$ relation, $m_2^{\prime}$, as
\begin{equation}
\beta = - \frac{0.4}{c_{2}^{\prime}}.
\end{equation}
As we have shown, the mean value of $c_{2}^{\prime}$ for the RS, 
within the high-confidence redshift range, is $-0.070\pm0.006$, before AC is applied. This value implies a L-$\sigma$ relation slope of $\beta=5.71 \pm 0.49$. 
Imposing a constant offset of $-0.019$ to the slope, as introduced by the AC according to Equation~\ref{eq:m2_corrected}, 
the aperture-corrected L-$\sigma$ relation becomes even steeper, with a mean slope of $\beta=7.83 \pm 1.11$. 

 The above measurement has been performed with a sample of more than $600,000$ massive LRGs, 
with a mean stellar mass of $M_* \simeq 10^{11.3} M_{\odot}$, within the redshift range $0.5<z<0.7$. 
At intermediate mass ranges and $z\sim0.1$, using ETG samples extracted from the SDSS, it has been 
shown that the slope of the L-$\sigma$ relation is consistent with that of a canonical F-J relation, i.e. $\sim 4$ 
(\citealt{Bernardi2003b, Desroches2007}). Even though this is a fairly robust result, studies at these mass ranges might 
still be subject to an inadequate treatment of selection effects. As an example, 
\cite{LaBarbera2010}, using the SDSS-UKIDSS survey, report a slope of $\sim 5$ for a sample at a similar mass range. 

At higher mass ranges, the curvature of the L-$\sigma$ relation has also been 
detected at high statistical significance at $z\sim0.1$ using the SDSS (\citealt{Desroches2007},
\citealt{Hyde2009a}, \citealt{Bernardi2011}). However, a definite quantification of the high-mass slope
 has not emerged. \cite{Desroches2007} report a slope of $\sim4.5$ at $M_r \geq -24$ (or $\log_{10} \sigma \gtrsim 2.4$). The authors find a much steeper 
slope, of $\sim5.9$, for a subsample of brightest cluster galaxies (BCGs).

Some other works have focused on small samples of ETGs in the nearby universe (of dozens to a couple of hundred objects). 
While these samples are strongly affected by low-number statistics, they have the advantage that galaxy properties 
can be measured with higher precision. \cite{Lauer2007b}, using a compilation of HST observations of a sample of 219 ETGs,
measure a slope for the L-$\sigma$ relation of $\sim7$ for a subsample of core and BCG ellipticals, the latter 
being predominantly core ellipticals as well (note that \citealt{Bernardi2007}, using an SDSS sample, report a significantly 
steeper slope for BCGs as compare to normal ETGs; from their Figure 6 we can visually estimate a slope of $\sim8$). Core ellipticals are identified by the 
fact that the central light profile is a shallow power law separated by a break from the outer, steep Sersic function profile. 
A value of $\sim2$ is reported for the coreless or "power-law" elliptical subsample, a class of objects that present steep cusps in surface brightness.
Interestingly, a number of studies suggest that core ellipticals dominate at the high-mass end, whereas coreless ellipticals 
are predominant at lower masses (see e.g. \citealt{Faber2007,Lauer2007a,Lauer2007b,Hyde2008}). Subsequently, \cite{Kormendy2013}, using a similar sample with 
some corrections and slight modifications in the core/coreless classification, estimate a slope of $\sim8$ for the core 
sample alone, and close to the canonical value of $4$ for the coreless sample. Our results for the slope of the 
L-$\sigma$ relation are in excellent agreement with these studies, in terms of core ellipticals.

 Results consistent with the above picture are also obtained from the ATLAS$^{3D}$ sample \citep{Cappellari2011}, comprising
 260 local-volume ETGs. \cite{Cappellari2013_XX} measure a high-mass slope of $\sim4.7$ (for the related stellar mass M-$\sigma$ relation). 
This result is obtained above a characteristic stellar mass of $\sim 2 \times 10^{11} M_{\odot}$, 
a range of masses where the ETG population in this sample is again reportedly dominated by core ellipticals. Note that
a detection of this mass scale in a {\it{statistically-significant}} sample came with the SDSS  \citep{Bernardi2011}. This scale corresponds approximately to the range of masses covered by BOSS.

Independently of the central brightness profile classification, a steep slope at the bright/massive end is confirmed by other studies. In the 
environment analysis of \cite{Focardi2012} (using a small sample of a few hundred objects extracted from HYPERLEDA, \citealt{Paturel2003}) a value of $\sim5.6$ is 
reported at high luminosities, although the authors indicate that an alternative method to compute this slope could yield a value closer to $4.5$.

To summarize, we have drawn two main conclusions in terms of the slope of the high-mass L-$\sigma$ relation 
from a study of previous literature at low redshift. Firstly, enough evidence has been gathered 
of a steeper slope at the high-mass end from SDSS works at $z\sim0.1$. Secondly, 
the reported value for the slope varies among different works, from $\sim4.5$ to $\sim8$. 
Note that the L-$\sigma$ relation is sensitive to low-number / selection effects 
(which nearby samples are affected by), but also 
to the region of the FP probed (in particular, to the exact luminosity range under analysis). Our work, 
provides, for the first time, a measurement at an unprecedented statistical significance of the slope at an intermediate redshift and at the highest-mass end, where the 
L-$\sigma$ relation``saturates" (using the terminology of \citealt{Kormendy2013}, among other authors).

Another important result from this work is the small intrinsic scatter of the L-$\sigma$ relation 
at the high-mass end at $z\sim0.55$; a measurement that has been performed with high statistical significance
for the first time. The scatter, quantified by the hyperparameter $s$ in this work, is found to have a mean value of $s=0.047 \pm 0.004$ in $\log_{10} \sigma$
at fixed L and redshift slice (with no significant redshift dependence). Again, in order to place these results
in the context of previous literature we can only compare with nearby/low-redshift samples, where some clear 
indications have been reported that the intrinsic scatter is smaller at higher masses. 

Figure 9 (right panel) from \cite{Hyde2009a} clearly shows that the
scatter in the L-$\sigma$ relation (in particular, in the $\log_{10} \sigma$ - $M_r$ relation) decreases towards high luminosities. 
In particular, we can visually estimate a value of $\sim0.05$ dex at $M_r < -23$.
A similar value for the scatter of the mass-$\sigma$ relation at the high-mass end can be visually estimated from Figure 1 
of \cite{Bernardi2011}. Although these values are not explicitly reported, they are obtained from 
relatively large SDSS samples (the \citealt{Bernardi2011} sample contains $\sim18,000$ massive ETGs), which 
indicates that these results are statistically significant. 
 
In nearby samples, however, low number statistics prevent a reliable estimation of the scatter. In any case, indications have 
been reported that the intrinsic scatter in smaller in core ellipticals, which would be consistent with our measurement.
At the high-mass end, \citet{Kormendy2013}, by adopting measurement errors of 0.1 mag in magnitude and 0.03 in $\log_{10} \sigma$, report an intrinsic physical scatter 
of 0.06 in $\log_{10} \sigma$ for core galaxies and 0.10 in $\log_{10} \sigma$ for core-less galaxies at a given magnitude.

Even though we must be careful to compare 
results at the high-mass end from a quantitative point of view, a mass (or luminosity) dependence of the intrinsic scatter of the L-$\sigma$ relation
have been reported in other works, including \cite{Sheth2003}, \cite{Desroches2007}, \cite{NigocheNetro2011} and \cite{Focardi2012}. 

 Finally, our result that the evolution of the zero-point of the L-$\sigma$ relation is consistent with 
a passive model with a formation redshift of $z=2-3$ is in good agreement with the redshift evolution of the FP as
measured in galaxy clusters up to $z\sim1$ (see e.g. \citealt{vanDokkum1996} at $z=0.39$; \citealt{Kelson1997} at $z=0.58$;
\citealt{vanDokkum1998} at $z=0.83$; \citealt{vanDokkum1998} at $z=0.83$; \citealt{Wuyts2004} at $z=0.583$ and $z=0.83$;
 \citealt{vanDokkum2003,Holden2005} at $z\sim1.25$). More generally, the result that the high-mass RS population evolves passively 
 from a high formation redshift is in agreement with a wide array of analyses (see, e.g., \citealt{Wake2006,Cool2008}, MD2014 for LF
 results; \citealt{Maraston2013} for LRG-SED evolution results; \citealt{Guo2013,Guo2014} for LRG-clustering results).

In light of the above comparison, it appears that the high-mass end of the L-$\sigma$ relation not 
only remains unchanged within the redshift range $0.5<z<0.7$, but would also be consistent 
with $z\sim0$ results.

\subsection{Physical interpretation} 
\label{sec:interpretation}

The very steep slope of the L-$\sigma$ relation at the high-mass end implies that the 
interplay between the different processes involved in shaping the evolution of RS galaxies at 
different mass ranges is systematically different. As galaxies grow, central velocity dispersions do not change as much 
as expected according to the scaling relations at lower masses. This result is 
consistent with the systematic variation of the total mass profile as a function of mass 
in the central region of ETGs found by \cite{Shu2015}, so that more massive galaxies
have shallower profiles. One possibility to explain this behavior is that the relative 
efficiencies of gas cooling and feedback in RS galaxies vary at different mass scales. 
Gas cooling permits baryons to condense in the central regions of galaxies, and therefore 
it is believed to make the mass distribution more centrally concentrated \citep[e.g.][]{Gnedin2004, Gustafsson2006, Abadi2010, Velliscig2014}. 
Heating due to dynamical friction and supernovae (SN)/Active Galactic Nucleus (AGN) feedback, in contrast, 
can soften the central density concentration \citep[e.g.][]{Nipoti2004, Romano-Diaz2008, Governato2010, Duffy2010, Martizzi2012, Dubois2013, Velliscig2014}. 
If feedback became more efficient in more massive galaxies, that could explain both 
the results presented here and in \cite{Shu2015}.

 A more straightforward explanation comes from a scenario where massive and intermediate-mass
ETGs have a different evolutionary history. From high-resolution images of a small number of local ETGs, some evidence has been gathered that 
the high-mass end of the RS population might be occupied almost exclusively by core ellipticals, 
whereas coreless ellipticals dominate at intermediate and lower masses \citep{Lauer2007a, Lauer2007b,Hyde2008,Cappellari2013_XX}. 
This transition occurs at $\log_{10}M_* \sim 11.2$, a mass scale that was first detected {\it{with high-statistical significance} by 
\cite{Bernardi2011}.}Even though this classification arises from the shape of the central surface brightness profile, several 
studies have shown that this bimodality, known as the ``E-E dichotomy", extends to a number of other galaxy properties: core ellipticals have boxy isophotes and 
are slow rotators while coreless ellipticals have more disky isophotes and rotate faster (to name but 
a few properties, see e.g. \citealt{Kormendy1996,Faber1997,Lauer2007a, Lauer2007b,Hyde2008,Cappellari2013_XX, Kormendy2013} for a complete
discussion). Importantly, the above mass scale has been associated, with high statistical
significance, with the curvature of the scaling relations (see \citealt{Hyde2009a, Bernardi2011}).

This characterization confirms the ideas of \citealt{Kormendy1996}, who proposed
a revision of the Hubble Sequence for elliptical galaxies, where isophote shape is used as an implicit indicator of velocity anisotropy.
The general consensus is that the properties of these two distinct populations are the consequence of
two different evolutionary paths. Massive core ellipticals appear to be formed through major dissipationless 
mergers \citep{Lauer2007a, Lauer2007b,Bernardi2011,Cappellari2013_XX, Kormendy2013}, whereas the less-massive coreless ellipticals seem to have 
undergone a more complex evolution (\citealt{Kormendy2009} review evidence that they are formed in wet mergers 
with starbursts).

 This paper provides the most precise measurements of the L-$\sigma$ relation at the highest mass
range ever probed with statistical significance. Unfortunately, however, BOSS does not provide the type of 
data required to perform a central surface brightness profile analysis that can 
confirm that the sample is dominated by core ellipticals, as previous results suggest. A complete analysis 
of this type would answer the question as to whether core ellipticals are fully responsible for 
the steep slope. This option is claimed by \cite{Kormendy2013} by showing that the slope of the 
L-$\sigma$ relation is significantly steeper for core galaxies even in the luminosity region where core and coreless
galaxies overlap (recall the statistical limitations of the study). Independently of this discussion, the shallow dependence
of $\sigma$ on galaxy mass (obtained from a L-$\sigma$ relation very similar to what we obtained) is reported in \cite{Kormendy2013} to be
approximately similar to N-body predictions \citep[][]{Nipoti2003, BoylanKolchin2006, Hilz2012} for dissipationless major mergers.

 Previous literature about the ``E-E dichotomy" along with the similarities between our measurements and those 
reported by \cite{Lauer2007a} and \cite{Kormendy2013} suggest that the intrinsic high-mass RS distribution characterized in MD2014
(the same distribution for which we compute the L-$\sigma$ relation here) can be identified as 
a core-elliptical population. In MD2014, the intrinsic RS distribution is photometrically 
deconvolved from photometric errors and selection effects. The resulting distribution is so narrow 
in the colour-colour plane that is consistent with a delta function, at fixed magnitude and narrow redshift slice (with 
a shallow colour-magnitude dependence for its location). The second component of the intrinsic model, the BC, is a more extended distribution, 
well described by a Gaussian function in the colour-colour plane, upon which the RS is superimposed. Our BC is defined
as a background distribution including everything not belonging to the very-pronounced RS (see MD2014 for details). 
This BC actually extends through the red side of the colour-colour plane. This characterization 
could reflect the ``E-E dichotomy" on the red side of the colour-colour plane, with this intrinsic BC being composed by a large fraction of coreless ellipticals, for which more 
scatter on the colour-colour plane is to be expected. Follow-up work will be needed to confirm this picture.

The intrinsic scatter of the L-$\sigma$ relation at $z=0.55$ adds to the result reported in MD2014 that the intrinsic RS distribution
is extremely concentrated in the colour-colour plane as well (at fixed magnitude and narrow redshift slice), 
which is an indication that there is little variability as far as the stellar populations are concerned within this population.
The small scatter in these relations is consistent with the idea that this is an aged quiescent population that has seen little activity in a long time.

The above idea is reinforced by the fact that the high-mass L-$\sigma$ relation at $z=0.55$ appears to be
very similar to that reported at $z=0.1$. Also, the evolution of the zero-point within the redshift range
$0.5<z<0.7$ is consistent with that of a passively-evolving population that formed at redshift of $z=2-3$, in agreement with 
the LF-evolution results shown in MD2014 for the same population. The picture of the evolution of the high-mass end of the RS is, however, 
not completely clarified yet. \cite{Bernardi2015} have recently analyzed the high-mass luminosity and stellar mass function 
evolution from the CMASS to the SDSS, reporting a puzzling result. The evolution of this functions appears to be ``impressively"  passive, when K+E corrections computed 
from the \cite{Maraston2009} models are used. However, when matched in comoving number- or luminosity-density, the SDSS galaxies 
are less strongly clustered than CMASS galaxies, which is obviously inconsistent with a passive evolution scenario.

\section{Conclusions and Future Applications} 
\label{sec:conclusions}

We have measured the intrinsic L-$\sigma$ relation for massive, luminous red galaxies
within the redshift range $0.5<z<0.7$. We achieve unprecedented precision at 
the high-mass end ($M_* \gtrsim 10^{11} M_{\odot}$) on the measurement of the parameters 
of the L-$\sigma$ relation by using a 
sample of 600,000 galaxies from the BOSS CMASS sample (SDSS-III). 
We have deconvolved  the effects of photometric and spectroscopic uncertainties and red--blue 
galaxy confusion using a novel hierarchical Bayesian formalism that is generally applicable to any 
combination of photometric and spectroscopic observables. The main conclusions of our analysis 
can be summarized as follows:

\begin{itemize}     
         \item At $z\sim0.55$, the passively-evolved L-$\sigma$ relation at $M_* \gtrsim 10^{11} M_{\odot}$ appears to be consistent with that at $z=0.1$.
	\item The slope of the $z=0.55$ L-$\sigma$ relation at the high-mass end is $\beta = 7.83 \pm 1.11$, 
	corresponding to the canonical form $L_i \propto \sigma^{\beta}$. 
	This value confirms, with the highest statistical significance ever achieved, the idea of a curved mass-dependent L-$\sigma$ relation. 
	Scaling relations for the most massive LRGs are systematically different than the relations defined at lower masses.	
	\item The intrinsic scatter on the L-$\sigma$ relation is $s=0.047 \pm 0.004$ in $\log_{10} \sigma$ at fixed L. This value confirms, with the highest 
	statistical significance ever achieved, that the intrinsic scatter decreases as a function of mass.	
	\item We detect no significant evolution in the slope and scatter of the L-$\sigma$ relation within the redshift range 
	considered. Under a single stellar population assumption, the redshift evolution of the zero-point is consistent within the errors with that of a
	passively-evolving galaxy population that formed at redshift $z=2-3$. This is in agreement with the LF-evolution results reported 
	in MD2014 for the same population. 	
	\item Our results, in combination with those reported in MD2014, provide an accurate 
	description of the high-mass end of the red sequence population at $z\sim0.55$, which is characterized in MD2014 as an extremely 
	narrow population in the optical colour-colour plane.	
	\item Our results for the L-$\sigma$ relation, in the light of previous literature, suggest that our high-mass RS
	distribution might be identified with the ``core-elliptical" galaxy population. In light of the ETG dichotomy, 
	the second component identified in MD2014, a much more extended distribution upon which the RS is superimposed, would 
	contain a significant fraction of ``coreless" ellipticals towards the red side. The larger scatter in colour found for this population 
	would be consistent with the evolutionary path that has been proposed for coreless ellipticals. 
\end{itemize}

 The above results lead us to consider followup work intended to investigate core--coreless elliptical 
demographics in BOSS. This project will require the use of high-resolution data. 

 The success of our algorithm for the photometric deconvolution of spectroscopic observables opens 
a field of future applications, as it can be used to constrain the intrinsic distribution of a variety of spectroscopically-derived 
quantities in BOSS. In the broader picture, the statistical techniques developed in this work and in MD2014 
lay the foundations for galaxy-evolution studies using other current and future dark energy surveys, like 
eBOSS, which are subject to the same type of SN limitations and selection effects that we face in BOSS.

 The extensive characterization of the high-mass RS presented in this work and in MD2014 will 
be used in combination with N-body numerical simulations to investigate the intrinsic clustering properties 
of this galaxy population, along with the intrinsic connection between these galaxies and the dark matter haloes that they inhabit.
The connection between galaxies and halos will be performed by applying the techniques of halo occupation distributions 
(HOD: e.g., \citealt{Berlind2002}; \citealt{Zehavi2005}) and halo abundance matching (HAM: e.g., \citealt{Vale2004}; \citealt{Trujillo2011}).
This is a novel approach as compared with the previous clustering/halo-galaxy-connection studies in BOSS, which have 
focused on the observed galaxy distribution and lacked a proper completeness correction.

 Finally, the velocity-dispersion distributions implied by the L-$\sigma$ relation that we 
have obtained in this work can be used in combination with the luminosity-function results of 
MD2014 to determine the statistical strong gravitational lensing cross section 
of the CMASS sample. This cross section can in turn be used to 
predict and interpret the incidence of spectroscopically selected strong lenses within large redshift 
surveys (e.g., \citealt{Bolton2008}, \citealt{Brownstein2012}, \citealt{Arneson2012}, 
\citealt{Bolton2012b}), and to derive constraints on cosmological parameters from the 
statistics of gravitationally lensed quasars (e.g., \citealt{Kochanek1996}, \citealt{Chae2002}, \citealt{Mitchell2005}).

\section*{Acknowledgments}

This material is based upon work supported by the U.S. Department of Energy, Office of Science, 
Office of High Energy Physics, under Award Number DE-SC0010331.

The support and resources from the Center for High Performance Computing at the University of Utah are gratefully acknowledged.

Funding for SDSS-III has been provided by the Alfred P. Sloan Foundation, the Participating Institutions, the National Science 
Foundation, and the U.S. Department of Energy Office of Science. The SDSS-III Web site is http://www.sdss3.org/.

SDSS-III is managed by the Astrophysical Research Consortium for the Participating Institutions of the SDSS-III Collaboration 
including the University of Arizona, the Brazilian Participation Group, Brookhaven National Laboratory, University of Cambridge,
University of Florida, the French Participation Group, the German Participation Group, the Instituto de Astrofisica de Canarias, 
the Michigan State/Notre Dame/JINA Participation Group, Johns Hopkins University, Lawrence Berkeley National Laboratory, Max Planck
Institute for Astrophysics, New Mexico State University, New York University, Ohio State University, Pennsylvania State University, University of 
Portsmouth, Princeton University, the Spanish Participation Group, University of Tokyo, The University of Utah, Vanderbilt University, University
of Virginia, University of Washington, and Yale University.

\bibliography{./paper}

\label{lastpage}

\end{document}